\def\kms{\ifmmode{\,\hbox{km}\,s^{-1}}\else {\rm\,km\,s$^{-1}$}\fi}

\def\msun{{\rm\,M_\odot}}
\def\lsun{{\rm\,L_\odot}}

\def\kmsm{{\rm\,km\,s^{-1}\,Mpc^{-1}}}

\def\hmpc{\ifmmode{h^{-1}\,\hbox{Mpc}}\else{$h^{-1}$\thinspace Mpc}\fi}
\def\hkpc{\ifmmode{\,h^{-1}\,{\rm kpc}}\else {$h^{-1}$\,kpc}\fi}

\def\et{{\it et~al.}~}

\def\ie{{\it i.e.}~}

\def\rpmax{r_p^{max}}
\def\rzmax{r_z^{max}}
\def\dvmax{\Delta v^{max}}

\documentclass[preprint]{aastex}
\tighten

\begin{document}

\title{Galaxy Groups at Intermediate Redshift}

\author{R.~G.~Carlberg\altaffilmark{1,2},
H.~K.~C.~Yee\altaffilmark{1,2},
S.~L.~Morris\altaffilmark{1,3},
H.~Lin\altaffilmark{1,2,4,5}, \\
P.~B.~Hall\altaffilmark{1,2},
D.~R.~Patton\altaffilmark{1,2},
M.~Sawicki\altaffilmark{1,2,7}, 
and
C.~W.~Shepherd\altaffilmark{1,2}
}
\altaffiltext{1}{Visiting Astronomer, Canada--France--Hawaii Telescope, 
        which is operated by the National Research Council of Canada,
        le Centre National de Recherche Scientifique, and the University 
	of Hawaii.}
\altaffiltext{2}{Department of Astronomy, University of Toronto, 
        Toronto ON, M5S~3H8 Canada}
\altaffiltext{3}{Dominion Astrophysical Observatory, 
        Herzberg Institute of Astrophysics,    ,  
        National Research Council of Canada,
        5071 West Saanich Road,
        Victoria, BC, V8X~4M6, Canada}
\altaffiltext{4}{Steward Observatory, University of Arizona,
        Tucson, AZ, 85721}
\altaffiltext{5}{Hubble Fellow}
\altaffiltext{6}{Department of Physics \& Astronomy,
        University of Victoria,
        Victoria, BC, V8W~3P6, Canada}
\altaffiltext{7}{Mail Code 320-47, Caltech, Pasadena 91125, USA}


\begin{abstract} 
Galaxy groups likely to be virialized are identified within the CNOC2
intermediate redshift galaxy survey using an iterative method. The
resulting groups have a median velocity dispersion of about 200
\kms. The virial mass-to-light ratios, using k-corrected and
evolution-compensated luminosities, have medians in the range of
$150-250h\msun/\lsun$. The number-velocity dispersion relation is in
agreement with the low-mass extrapolation of the cluster normalized
Press-Schechter function.  The two-point group-group correlation
function has $r_0=6.8\pm 0.3\hmpc$, which is larger than the
correlations of individual galaxies at the level predicted from n-body
calibrated halo clustering.  We conclude that the global statistics of
groups are in approximate accord with dark matter halo
predictions. The groups are stacked in velocity and position to create
a sample large enough for measurement of a density and velocity
dispersion profile.  The resulting stacked group contains about 1000
members above a well defined background distribution. The stacked mean
galaxy density profile falls nearly as a power law with $r^{-2.5}$ and
has no well-defined core.  The projected velocity dispersion is
examined for a variety of samples with different methods and found to
be either flat or slowly rising outwards. The combination of a
steeper-than-isothermal density profile and the outward rising
velocity dispersion implies that the mass-to-light ratio of groups
rises with radius. The $M/L$ can be kept nearly constant if the galaxy
orbits are nearly circular, although such strong tangential anisotropy
is not supported by other evidence.  The segregation of mass and light
is not dependent on galaxy luminosity but is far more prominent in the
red galaxies than the blue.  The $M/L$ gradient could arise from
orbital ``sloshing'' of the galaxies in the group halos, dynamical
friction acting on the galaxies in a background of ``classical''
collisionless dark matter, or, more speculatively, the dark matter may
have a true core.
\end{abstract}


\section{Introduction}

Small groups of galaxies are important cosmological indicators of the
distribution and properties of the dark matter in the universe.  They
occupy the mass and velocity range between individual galactic halos
and the large halos of rich galaxy clusters
\citep{abellcat,bb,gott+turner,hickson,cfa_xi,nw}.  The RMS velocity
dispersion of groups is only somewhat larger than that of individual
galaxies, but groups have the advantage that visible galaxy tracers
extend throughout the dark matter halo. At intermediate redshift,
groups are suitable targets for X-ray observation and weak
gravitational lensing studies which are complementary probes of their
contents..  Consequently groups can be used to probe the properties of
the dark matter on scales and at velocities much smaller than can be
examined in galaxy clusters.

The theory of structure growth in the universe is based on the
paradigm that the dark matter consists of collisionless particles that
only interact via the gravitational force. Cold Dark Matter is a
specific form of this hypothesis that has been subjected to intensive
theoretical study.  A particular strength is that the properties of
virialized halos can be predicted from a given density perturbation
spectrum to full nonlinearity via simulations and various analytic
approximations. These predictions have been tested with varying
degrees of success against the dark matter halos of individual
galaxies and rich clusters, but are less examined on intermediate
scales.  The intermediate scales are interesting because they are at
much higher phase space densities than massive galaxy clusters, yet
their central dark matter densities are not overwhelmed and altered by
the baryonic matter, as is the case in for most normal galaxies.

The CDM theory gives specific predictions of the statistical
properties of the dark halo population and the mean internal
properties of individual halos.  The Press-Schechter theory (1974,
hereafter PS) predicts the numbers of halos as a function of mass or
velocity dispersion. At low redshift, suitably selected groups have a
population volume density in accord with the cluster-normalized PS
prediction \citep{moore,girardi}. A second global statistic is the
clustering of dark matter halos which is predicted using analytic
approximations which have been compared to n-body results
\citep{mowhite,jing}. This biased ``peaks'' theory predicts a slow increase of
clustering strength with halo mass.  The internal density structure of
the halos is found in simulations to have a power law cusp, $r^{-1.0}$
to $r^{-1.5}$ which asymptotically steepens to approximately $r^{-3}$
beyond the virial radius \citep{dc,nfw,moore_core,klypin_gp}. 
The goals of this paper are to find a collection of virialized groups
and then to compare the predictions of the global statistics and
internal structure of intermediate mass dark matter halos to the
observationally derived properties.  

The paper is organized as follows. The next section describes our
approach to identifying groups in a redshift catalogue.  Section 3
gives an overview of the Canadian Network for Observational
Cosmology's field galaxy redshift survey (CNOC2) and the virialized
groups that we find.  In \S4 the number-velocity dispersion relation
and the two-point correlation function are computed and compared to
dark matter halo predictions. In
\S5 we ``stack'' the groups on their centers and measure the mean
projected density distribution and projected velocity distribution.
In \S6 we model these as projections of simple 3D systems to derive
the mean mass density profile and mass-to-light ratio as a function of
radius which leads to the discovery that groups have a rising
mass-to-light ratio. We examine the rising mass-to-light ratio for
various galaxy sub-populations to search for systematic trends that
might point to its physical origin. We conclude with a discussion of
the possible implications of these results and a short set of
empirical conclusions.  We use $H_0=100h \kmsm$ throughout this paper
and adopt $\Omega_M=0.2, \Omega_\Lambda=0$ as our reference
cosmological model. The distances and transverse lengths would be
about 8\% larger in an $\Omega_M=0.3, \Lambda=0.7$ cosmology at the
median redshift.

\section{Finding Groups in Redshift Space}

A group is defined here as a collection of three or more galaxies,
above some minimum luminosity, that meets a set of positional
requirements designed to minimize chance associations.  One might
seek, say, bound groups, such as the low overdensity Local Group, or,
virialized groups, which are collapsed and hence quite
dense. Virialized groups are a high density subset of bound groups and
are the aggregates of interest in this paper.  Unfortunately, in
redshift space there is a fundamental degeneracy between position and
line of sight velocity. Consequently, the precise galaxy membership of
a group found in redshift space is always a statistical issue.

Most group search methods are based on the friends-of-friends (f-o-f)
algorithm used by Huchra \& Geller (1982). This is an important method
that gives unique groups, independent of starting galaxy. The f-o-f
algorithm can be tuned to yield groups of varying overdensity.  The
algorithm starts with any galaxy as the beginning of a trial
group. All galaxies closer than some maximum distance (discussed
below) are added to the group.  Then, each of the new group members is
in turn used as a center to search for its neighbors to add to the
growing group, continuing until no more new neighbors are found. Then
one proceeds to a previously un-examined galaxy to try to start a new
group. This process continues until all galaxies have been examined
for neighbors. Groups of one and two are then deleted from the
catalogue.

The f-o-f algorithm has two parameters in redshift space: a maximum
separation in projected radius, $\rpmax$, and either a maximum
separation velocity, $\dvmax$, or co-moving redshift space distance
difference, $\rzmax$, required to join the group. The $\dvmax$ and
$\rzmax$ parameters are related through $\Delta v= H(z)r_z/(1+z)$. We
use $\Delta v$ for kinematic measurements and $r_z$ for group finding.
The $\rpmax$ and $\rzmax$ parameters need to be mutually adjusted to
take into account the mean volume density of the survey, $n(z)$, so as
to produce an overdensity with respect to the field chosen on the
basis of an experimental goal with an allowance for redshift space
blurring.  The resulting galaxy overdensity in the cylindrical
redshift space search volume is,
\begin{equation}
{\delta n\over n_0} \simeq {1\over{2\pi[ \rpmax]^2 \,\rzmax \,n(z)}}.
\label{eq:od}
\end{equation}
The $\rzmax$ parameter needs to be chosen in relation to $\rpmax$ with
some care. One approach is to tune it to n-body simulation results
\citep{nw}.  However, an ever-present issue is that the resulting 
group sizes and velocity dispersions tend to correlate with the
$\rpmax$ and $\rzmax$ parameters.

Here we are interested exclusively in virialized groups, suggesting we
devise a variant of the basic f-o-f algorithm which selects groups
that appear to be sufficiently dense that they will quickly virialize.
In configuration space virialization demands a mean interior density
of approximately $178\Omega^{0.45}\rho_c$
\citep{eke_etal}, equivalent to nearly $350\rho_0$ in a low density,
flat cosmology.  A conventional approximation is that such groups will
be contained inside the radius $r_{200}$ which can be estimated from
the virial theorem as $r_{200}=\sqrt{3} \sigma_1/[10 H(z)]$, where
$\sigma_1$ is the line-of-sight velocity dispersion and $H(z)$ is the
Hubble constant at the redshift of interest \citep{cnoc1}. For a given
$\sigma_1$ the virialized group members will be, on the average,
contained with approximately $1.5 r_{200}$ if the mean density is
falling like $r^{-3}$. This relation between velocity and projected
separation immediately suggests a natural range for the
parameters. The field pairwise velocity dispersion is approximately
300 \kms\ \citep{dp}, equivalent to a single galaxy random velocity of
about 200 \kms. This velocity dispersion is generated in groups which
suggests that an average $r_{200}$ will be about 0.3\hmpc\ at $z\sim
0.4$.

The f-o-f algorithm provides a set of trial groups whose properties
are fixed by the input link distance parameters such that many of them
may not be virialized. Virialized groups have a minimum overdensity of
about $200\rho_c$. Hence, for each trial group we estimate a velocity
dispersion which is then used to calculate $r_{200}$.  Galaxies beyond
a distance to the group center of $1.5 r_{200}$ are discarded. The
remaining galaxies are used to recalculate the velocity dispersion.
This can be iterated until the group converges to a stable set. Some
trial groups quickly drop to only one or two members and hence no
longer qualify for group status.  On the other hand, if we choose a
very large starting value of $\rpmax$ or $\dvmax$ a few groups will
percolate over very large structures. A minor complication is that we
must identify a group center.

The details of our group finding algorithm follow.  (1) Pick a
cosmology for the analysis ($H_0= 100 \kmsm, \Omega_M=0.2,
\Omega_\Lambda=0$). (2) Set the sample's redshift and absolute
luminosity limits (k-corrected and evolution-compensated at a mean
rate of one magnitude per unit redshift) of $M_R^{ke}=-18.5$ mag, no
initial redshift limits) which defines a galaxy sample for all further
operations. (3) Pick an $\rpmax$ (our standard groups use 0.25\hmpc)
and $\rzmax$ (5 \hmpc\ for our standard groups). Center a cylinder of
radius $\rpmax$ and forward and backward extent of $\rzmax$ on each
sample galaxy and count the number of sample galaxies.  To create a
background estimate we randomly draw points from an $n(z)$ fitted to
the full sample and count the number within the sample cylinder.  If
the search radii initially give less than three neighbors, then
multiply smoothing lengths by 1.5 and repeat for this sample galaxy.
At this stage a new subsample is defined requiring that the local
overdensity relative to the smooth sample have some specified minimum
value. In this paper we only impose the requirement that the local
overdensity be positive.  (4) Select the highest density ungrouped
galaxy and begin to find a new group. (5) Standing on each new group
member in turn, add to the group any galaxy in the minimum overdensity
subsample that is closer than $\rpmax$ and $\rzmax$. Repeat this step
until no new galaxies are added. (6) This f-o-f group defines the
starting group for the virialized group iteration.  (7) For the trial
virialized group, determine the geometric selection function weighted
mean $x$, $y$, $z$ and $\sigma_1$.  Trim galaxies beyond, or add
galaxies from the f-o-f list within, $r_p=1.5r_{200}$ and $\Delta v=
3\sigma_1$.  Repeat step 7 four more times, requiring that the last
two iterations have an identical result. (8) Drop single galaxies and
pairs from the catalogue.

\section{Groups in the CNOC2 Survey}

The Canadian Network for Observational Cosmology Field Galaxy Redshift
Survey (CNOC2) was undertaken primarily to study the dynamics of
galaxy clustering at intermediate redshift.  The survey methods and
catalogues are fully described in Yee \et\ (2000). The survey covers a
total of about 1.5 square degrees in four patches spread around the
sky for observing efficiency and to control cosmic variance. Galaxy
redshifts are obtained over the redshift range 0 to 0.7, with the
unbiased spectroscopic sample extending between redshifts 0.1 and
0.55. The catalogues contain approximately 6000 galaxy redshifts with
an accuracy between 70 and 100\kms, along with UBVRI photometry. The
groups are constructed from these catalogued galaxies. The luminosity
and clustering evolution of the sample as a whole has been previously
discussed \citep{huan_lf,cnoc2_xi}.

The CNOC2 sample has $m_R \le 21.5$ mag and well defined spectroscopic
completeness weights.  The average redshift completeness is about
45\%, with nearly 100\% completeness 3 magnitudes above the limit and
about 20\% at the limit. Within our primary redshift range, 0.10 to
0.55, the redshift completeness to the flux limit is higher than for
the sample as a whole.  From our luminosity functions
\citep{huan_lf} we calculate the probability that we will successfully
obtain a redshift as about 75\% with the other 25\% being for galaxies
at redshifts out of our primary redshift range
\citep{cnoc2_tech}. Together these imply that about 60\% of the
galaxies above the flux limit in the redshift range 0.1 to 0.55 have
measured redshifts.  If a group contains 3 galaxies within the survey
limits, then the probability from the cumulative binomial distribution
that we will obtain 3 redshifts is $(0.6)^3$ or 0.216.  If the group
contains 4 eligible galaxies then the probability we will obtain 3 or
more redshifts rises to 0.47. The probabilities of obtaining 3 or more
redshifts in groups of 5, 6, 7 and 8 eligible members are 0.68, 0.81,
0.90 and 0.95, respectively, assuming no complications from geometric
selection \citep{cnoc2_tech}.  Our average group contains 3.8 galaxies
with redshifts. We conclude that our roughly one-in-two sampling
allows us to detect about half of the three or more members groups
that are present. This level of completeness has no bearing on most of
our analysis so we normally do not attempt to compensate for this
effect.

There is a small effect due to groups at the boundaries of the
surveyed region. Examining a typical field we find that no more than
20\% of the groups have any of their $r_{200}$ area beyond the
boundary and no more than half of that. Consequently approximately
10\% of the group members are ``missing'', which will have little
effect on the velocity dispersions, and cause roughly a 10\%
diminution of the group masses, through the virial radius, and a
similar effect on the total light. In the presence of much larger
random fluctuations due to the small numbers in the groups this is not
a major concern.

The only two parameters that turn out to have much of an impact on the
groups are $\rpmax$ and $\rzmax$.  Our density requirement forces the
``raw'' groups to have a velocity dispersion strongly correlated with
the search distance in the redshift direction, $\rzmax$. The effect of
the iteration on the group velocity dispersion is shown in
Figure~\ref{fig:sig12}. The initial velocity dispersion is calculated
for the f-o-f groups and is hence very strongly correlated with the
chosen $\rzmax$. The iteration to find virialized groups allows the
$\sigma_1$ values to relax to more appropriate values. However some
less than ideal groups do persist which will lead us to continually
consider alternate samples throughout this paper.  Most of the groups
converge to a stable membership in one or two iterations, with only
1-2\% discarded due to failure to converge in four iterations.

Figure~\ref{fig:nz} shows the redshift distribution of the groups for
four group catalogues with increasing $\rzmax$. Beyond redshift 0.45
the sample becomes incomplete as the flux limit of $m_R=21.5$ mag
causes galaxies to slip below the sample absolute magnitude limit
$M_R^{ke}=-18.5$ mag. Both this figure and a detailed examination of
the lists of groups shows that the sets of groups have a large
overlap, independent of the search parameters.  That is, this
indirectly indicates that the set of group centers is relatively
insensitive to the group finding procedure. In the next section we
examine a variety of statistics to form a basis to select some group
catalogues as best suited to various analyses.

\section{Global Properties of Groups and Halos}

We will use three global properties of the groups to assess the degree
of correspondence of various group catalogues to dark halos.  These
are the mass-to-light ratio distribution, the abundance as a function
of velocity dispersion and the clustering properties. The virial mass
to R band light ratio, $M_{VT}/L_R$, of groups is an indicator of the
value of $\Omega_M$.  The number density of dark matter halos as a
function of their one-dimensional RMS velocity dispersion,
$n(\sigma_1)$, is an important test of the CDM clustering spectrum but
here is used as a guide to whether the number of high velocity
dispersion groups is reasonable.  The group-group auto-correlation as
a function of mean separation, $\xi_{hh}(r)$, is a test of clustering
theory. Together these indicators provide valuable information as to
whether galaxy groups have approximately the properties expected on
the basis that the galaxies are orbiting in a dark-matter dominated
potential that is at least partially virialized.

Our statistical goals for group selection are to maximize the number
of real groups and minimize the number of groups which contain
redshift space interlopers. The derived group attributes come in three
categories, with increasing sensitivity to the search
parameters. First is the group centers, $x, y$ and $z$; second are the
group extensions in $r_p$, and velocity, and third there are the
specific galaxies in the groups. For all purposes the locations are
key, whereas the extensions are secondary.  The extensions do enter
scaling relations, and the precise group membership is not relevant
for averaged, background subtracted measurements.

\subsection{Virial Mass-to-Light Ratio}

The ratio of the virial mass to the total luminosity, $L_R^{k,e}$, is
a valuable indicator of the CDM mass density of the universe.  It does
not measure any component of the mass that clusters weakly. The virial
mass-to-light ratio of groups has quite considerable scatter simply as
a result of both small number statistics and orbital projection. On
the basis of dynamical simulations Heisler, Tremaine \& Bahcall (1985)
found the dispersion in virial mass estimates as a result of these
fluctuations is nearly a factor of two above and below the true mass,
but this was comparable to other mass estimators.  We calculate the
virial mass, $M_{VT}$, using the galaxies in the groups with $r_p\le
1.5r_{200}$ and $\Delta v
\le 3\sigma_1$, following the CNOC methods \citep{cnoc1}. That is, 
\begin{equation} 
M_{VT} = {3\pi\over 2G}\sigma_1^2 R_h,
\label{eq:vt}
\end{equation} 
where the virial radius is evaluated for the galaxies identified as
being group members. The circularly averaged harmonic radius is
\begin{equation}
R_h^{-1}=\left(\sum_i w_i\right)^{-2} \sum_{i<j}{w_iw_j 
        {2\over{\pi(r_i+r_j)}} K(k_{ij})},
\label{eq:rh}
\end{equation}
where $k_{ij}^2=4r_ir_j/[(r_i+r_j)^2+s^2]$ and $K(k)$ is the complete
elliptic integral of the first kind in Legendre's notation
\citep{nr}. The softening, $s=2$ arcsecond,
eliminates the divergence for galaxies at the same radii from the
group center \citep{cnoc1}.  The luminosity, $L_R^{k,e}$, is
k-corrected, evolution compensated, and includes an extrapolation of
the luminosity function to allow for galaxies below the redshift
dependent absolute magnitude cutoff. The evolution is taken to be at a
mean rate of one magnitude per unit redshift.

If group galaxies are drawn from a universal luminosity function and
the ratio of dark mass to luminous mass is a constant then the median
$M_{VT}/L$ should be constant. The spread of the distribution of
values can be used as an indicator of the statistical reliability of
the group selection procedure.  In Figure~\ref{fig:mls} we plot the
median $M_{VT}/L$ against the fractional difference between the first
and third quartile $M_{VT}/L$ values. In the same system clusters have
$M_{VT}/L=380\pm 70 \msun/\lsun$ \citep{cnoc1}, where we have removed
the CNOC1 correction for the mean flattening of clusters.  The median
$M_{VT}/L$ increases with both $\rpmax$ and $\rzmax$.  Groups selected
with $\rpmax=0.5\hmpc$ and large $\rzmax$ have huge median $M/L$
values and a large spread between first and third quartile values.
Smaller $\rzmax$ lead to an increase in the spread and decrease in the
mean $M_{VT}/L$. The origin of this decrease in $M_{VT}/L$ with size
is at least partially a result of the internal $M/L$ gradient within
groups that we discuss below. Overall, the $\rpmax=0.25\hmpc$ groups
with $\rzmax \le 7\hmpc$ have the desirable property that both the
spread and the median of the distribution do not change too much with
$\rzmax$.

The derived properties of the groups have very substantial
uncertainties because of the small number of galaxies with
velocities. The errors in the velocity dispersion and the resulting
correlation with the derived virial mass-to-light ratios are
illustrated in Figure~\ref{fig:ml_sig} for the $\rpmax=0.25\hmpc$ and
$\rzmax=5\hmpc$ groups. The large errors are the dominant source of
the very strong correlation between the velocity dispersion and the
derived $M/L$, as a consequence of $M_{VT}$ as $3G^{-1}\sigma_1^2
r_v$, which accurately predicts the slope of the correlation visible
in the figure. The inset diagram restricts the sample to groups with
at least six members.  In this case there is some support for the
indication that the the mass-to-light ratio rises with velocity
dispersion.

\subsection{The Number Density-Velocity Dispersion Relation}

The number of groups as a function of their line-of-sight RMS velocity
dispersion is given in Figure~\ref{fig:lnsig} for a range of group
search parameters. The median velocity dispersions for
$\rpmax=0.25\hmpc$ are 192, 229, 256 and 266 for $r_z^{max}$ of 3, 5,
7, and 10\hmpc, respectively.  Below 100\kms\ is the regime of
individual galaxies, which reduces the number of groups. A small
effect is that the velocity precision of the survey is about 100\kms,
which artificially reduces the numbers of low velocity dispersion
groups. At this stage we recall that large values of the search length
in the redshift direction tends to include enough outlier galaxies
that a few groups are promoted into the high $\sigma_1$ tail of the
distribution. Given that high velocity dispersion groups also tend to
contain the most galaxies (for a constant $M/L$), large groups are the
easiest to find. Thus low membership groups with high velocity
dispersions are likely to have erroneously large $\sigma_1$ values.

The Press-Schechter (1974) theory works well to describe the abundance
of halos in n-body experiments and a range of observational data,
including clusters over a range of redshifts \citep{cnoc_s8} and low
redshift groups \citep{girardi}.  We can compare the Press-Schechter
prediction for the number of groups to our observations.  Because this
requires absolute numbers we will use the redshift range of greatest
completeness, roughly 0.2 to 0.45.

To calculate the expected density we follow the procedures outlined in
Carlberg et al. (1997) using $\Omega_M=0.2,
\Omega_\Lambda=0$ and $\sigma_8=1.0$. We calculate the mass-velocity
dispersion relation as
\begin{equation}
M_{1.5} = 8.6\times 10^8 \sigma_1^3~ {\rm s}^3\, {\rm km}^{-3} \msun,
\label{eq:psmass}
\end{equation}
where $\sigma_1$ is given in units of \kms\ and the mass is the
nominal value inside a virialized 1.5\hmpc\ sphere, as is appropriate
for rich clusters, to which we want to normalize these predictions.
The equivalent top-hat radius that contains this mass at the mean
density is
\begin{equation}
R_L \simeq 8.43 \Omega_z^{0.2p/(3-p)} \left[{{M_{1.5}}\over
{6.97\times10^{14}\Omega_M h^{-1} \msun}}\right]^{1/(3-p)} (1+z)^{-p/(3-p)}~\hmpc,
\label{eq:psrl}
\end{equation}
where $p\simeq 0.64$ is the rate of increase of mass with radius
\citep{wef,cnoc_s8} and $\Omega_z$ is the value of $\Omega_M$ at redshift $z$.
We evaluate the Press-Schechter relation as
\begin{equation}
n(M(\sigma_1))dM = {{-3\delta_c(z)}\over{(2\pi r_L^2)^{3/2}\Delta}}
         {{d\ln{\Delta}}\over{d\,M}} \exp{[-\delta_c^2(z)/2\Delta^2]}
        \,dM,
\label{eq:ps}
\end{equation}
where $\delta_c(z)=0.15(12\pi)^{2/3}\Omega^{0.0185}/D(z,\Omega)$,
\citep{nfw} and $\Delta(r_L)$ is the top-hat fractional linear mass variance in spheres
of radius $r_L$ calculated using a fitted CDM spectrum \citep{ebw}.
To determine the number density in bins of velocity dispersion we
simply integrate over the relevant range of masses. Note that the
normalization we have used automatically means that our group number
densities will match on to the CNOC1 clusters
\citep{cnoc_s8,borgani,girardi}.

The Press-Schechter predictions of number density for a median
redshift of 0.36 are displayed in Figure~\ref{fig:lnsig}.  The
subsample is contained in a volume of $1.8 \times 10^5 h^{-3} {\rm
Mpc}^3$ (or about 50\% more for a flat cosmology). Below 100 \kms\ the
sample is missing many halos for two reasons. First, individual galaxy
halos make up the majority of the halos in this regime. The velocity
dispersion of an $M_\ast$ elliptical is about equal to that of our
median group (which in itself suggests an evolutionary connection).
Second, because our velocity accuracy is about 100 \kms\ low velocity
halos are scattered into the next higher bin. At this stage it we
recall that our expected completeness rate for higher velocity
dispersion groups is about 50\% and even lower for those with velocity
dispersions comparable to individual galaxies.

The Press-Schechter predictions are in reasonable agreement with the
groups for $\rzmax=3\hmpc$ and $5\hmpc$, bearing in mind that the
random errors are at least $\sqrt{N}$. Smaller values of $\rzmax$ miss
high $\sigma_1$ groups, while larger values, $\rzmax \ge 7\hmpc$, produce
a few highly improbable groups with the velocity dispersions of rich
clusters.

There are three reasons to select the $\rpmax=0.25\hmpc$ and
$\rzmax=5\hmpc$ groups as the best suited to our analysis of
virialized halos with velocity dispersions of approximately 100-300
\kms, although other group selections parameters give rise to samples
that show a very similar set of $x, y, z$ locations.  The selected
groups have a relatively low dispersion in their $M/L$ values, their
$n(\sigma_1)$ distribution is close to the Press-Schechter prediction
with few high velocity outliers, and the dynamical analysis below
finds that our chosen redshift distance inclusion length pulls in most
of the group members so that the derived velocity dispersions are
fairly stable against the addition of more outlying members with
increasing cutoff velocity.

The locations on the sky of the ``standard'' $\rzmax=5\hmpc$ and
$\rpmax=0.25\hmpc$ groups for one of the patches are shown as the
points in Figure~\ref{fig:xy}. The circles indicate the $r_{200}$
radii. Note that some groups are quite compact with respect to this
radius. The parameters of all of the standard groups are given in
Table~1. The columns give the location with co-ordinates measured
relative to the designated group centers \citep{cnoc2_tech},
$\sigma_1$ (from which $r_{200}$ is calculated), the virial
mass-to-light ratio, the number of group members with redshifts, and
the mass. The random errors of the derived quantities, as estimated
using the Jackknife technique, are very large for most of the groups,
a straightforward consequence of the small numbers of members.

\subsection{The Two-point Group-Group Correlation Function}

A fundamental prediction of hierarchical dark matter clustering is
that clustering, as measured by the two-point group correlation
function $\xi_{GG}(r)$, should increase with the mass or velocity
dispersion of the halo \citep{kaiser84,WDEF}. Here we measure both the
redshift space correlation, $\xi(s)$,and the projected correlation
function, $w_p(r_p)$, both of which provide an indication of the
correlation length, $r_0$. The co-moving redshift space separation is,
\begin{equation}
s^2=[r(\onehalf z_1+\onehalf z_2)(\theta_1-\theta_2)]^2 + [r(z_1)-r(z_2)]^2,
\end{equation}
with $r(z)$ being the co-moving distance at redshift $z$. At
separations small compared to the pairwise-velocity converted to a
distance, $\sigma_{12}/H(z)$, $\xi(s)$ ceases to increase with
decreasing separation as the random velocities begin to dominate the
redshift space separation.  On larger scales $\xi(s)$ is expected to
be enhanced relative to $\xi(r)$ as a result of the ``compression
effect'' of systematic infall \citep{kaiser_zspace}.  

The projected correlation function, $w_p(r_p) = \int
\xi[\sqrt{r_p^2+r_z^2}]\, dr_z$, has the advantage that the
peculiar velocities have no effect on the result \citep{dp}.  We use
the classical $DD/DR-1$ estimator \citep{lss} which is suitable in the
strong correlation regime. The random distribution is derived from a
fit to the observed $n(z)$ distribution of the groups.  Pairs are
included for $r_z$ up to 30\hmpc, co-moving. For a power law
correlation function, $\xi(r)=(r_0/r)^\gamma$, the reduced projected
correlation function, $w_p/r_p$ is equal to $A(\gamma)\xi(r)$, where
$A(\gamma)=\Gamma(\onehalf)\Gamma(\onehalf(\gamma-1))/\Gamma(\onehalf
\gamma)$, a factor of 3.68 for $\gamma=1.8$ \citep{dp}.

We evaluate the correlations using the same procedures and programs
used in Carlberg et al. (2000) to measure the correlation of
galaxies.  Using the standard group sample ($\rpmax=0.25\hmpc$ and
$\rzmax=5\hmpc$) we evaluate $\xi(s)$ and $w_p(r_p)/r_p$ over the
redshift range 0.15 to 0.55. We use as much redshift range as possible
to boost the sample size. The resulting correlation functions are
displayed in Figure~\ref{fig:gpxi}. In the form plotted both functions
are dimensionless functions. Note that the upward offset of
$w_p(r_p)/r_p$ relative to $\xi(s)$ is a natural result of the
$A(\gamma)$ factor. The error flags displayed in the figure are the
square root of the number of galaxy pairs in each bin. Fitting the
measured redshift space correlations to the function
$\xi(r)=(s_0/r)^{1.8}$, gives $s_0=6.8\pm 0.3\hmpc$. The fit for the
projected correlation function finds $r_0=6.5\pm0.3\hmpc$. These
errors are formal fitting errors and do not include an allowance for
the patch-to-patch variance which likely dominates the random error.

The correlations measured for the CNOC2 galaxies over this redshift
range in this cosmology have a mean of $r_0=4.2\pm 0.2\hmpc$ where
this error includes the patch-to-patch variance
\citep{cnoc2_xi}.  The ratio of the correlation amplitude of our
groups to that of the galaxies is $2.2\pm0.4$. At low redshift,
Ramella, Geller \& Huchra (1990) found $s_0\simeq8\hmpc$ for the CfA
groups, which are somewhat lower mean internal density than ours. A
similar result emerged for the combination of the CfA and SSRS2 groups
\citep{girardi_xi} which found that the group 
correlation amplitude was a factor of $1.64\pm0.16$ stronger than that
of galaxies, although the CfA groups alone appeared to have a smaller
offset.

The biasing theory of Mo \& White (1996) gives predictions of the
expected ratio of correlation amplitudes.  The characteristic velocity
dispersion of normal galaxies is about 100 \kms, whereas our groups
have a median velocity dispersion of about 200
\kms. From the CDM power spectrum
\citep{ebw} with a shape parameter of $\Gamma=0.2$, we calculate the
linear mass variances to be 3.3 and 2.4, for 1 and 2\hmpc\ top-hat
perturbations, which are approximately the unperturbed mean radii of
the perturbations associated with galaxies and groups,
respectively. With $\delta_c\simeq 1.68$ in the Mo \& White model, we
find that groups should be more strongly correlated by a factor of
about 1.75 or 1.47 with the Jing (1998) n-body calibrated
modification. Provided that the group velocity dispersion is about
twice the galaxy velocity dispersion the results are not very
sensitive to the velocity dispersions chosen.  We conclude that the
correlation amplitudes are consistent with the expected relation at
about the 1.5 standard deviation level.

At this stage we have shown that our sample of groups is in good
accord with three global statistics predicted on the basis of
galaxies tracing cold dark matter halos. The next issue is to examine
the relative internal distribution of the dark matter relative to the
galaxies.

\section{Mean Internal Structure of the Groups}

N-body simulations now have sufficient length resolution that they can
reliably predict the highly nonlinear realization of the collapse
of halos and their resulting internal density profile. A fundamental
prediction is that CDM halos have a central density cusp, roughly
$r^{-1.0}$ to $r^{-1.5}$
\citep{dc,nfw,moore_core,klypin_gp}. At large
radii the density begins to drop as approximately $r^{-3}$. The
characteristic radius of the density profile can be calibrated in
n-body experiments and derived from an approximate theory \citep{nfw}.
For the massive dark matter halos of rich clusters the NFW density
function, $\rho(r)=Ar^{-1} (r+a)^{-2}$, is entirely consistent with
the derived mass distribution
\citep{cnoc_nfw}. The situation for galaxy mass halos is somewhat
controversial, with constraining data coming from disk rotation curves
in the presence of substantial amounts of baryonic mass and the
modeling complication of partial pressure support relative to the
circular velocity. However there is evidence that in small velocity
dispersion halos CDM may allow central densities that exceed the
observations \citep{moore_core,moore_sub}. 

A fundamental difference between a galaxy and a rich cluster of
galaxies is that galaxies have a strongly rising mass-to-light ratios
with increasing distance from the center whereas rich galaxy clusters
have a nearly constant mass-to-light ratio over their virialized
volume \citep{cnoc1_pro}.  Therefore a basic question is whether
groups exhibit a rising mass-to-light ratio. Individual groups have
too few galaxies to make such a measurement. Moreover groups come with
a wide range of galaxy contents and have somewhat uncertain
virialization because of their small numbers. Therefore we assemble a
``stacked'' mean group to boost the numbers to levels where we can
make reliable measurements of the density and velocity dispersion
profiles.

\subsection{The Stacked Mean Group}

Precisely how the mean group is built from the individual groups will
have a significant impact on its properties. For instance, poorly
determined centers will create a core in the projected profile, or,
overlaying a small low velocity dispersion group with a large high
velocity dispersion group will lead to a rising velocity dispersion
with radius.  The idea is simple: we stack galaxies in both $r_p$ and
$\Delta v$ on the group centers keeping track of the expected number
of background galaxies.  All galaxies are used, with no distinction
given to galaxies that were used to define the group center.  The
group centers are determined in the virial analysis as the geometric
weighted mean of the locations of the iteratively selected group
members.  We have adopted geometric weights, which help to compensate
for the objects without redshifts. The smoothing radius used to
calculate our geometric weights is 120\arcsec\ which is so large that
there is little weight variation within these small groups, but
weights do vary from group to group. That is, at the $x,y,z$ location
of each of the $n_G$ groups in turn, we count the number of galaxies
in the neighborhood at separations $r_p/r_{200}$ and $\Delta
v/\sigma_1$, to measure the stacked density distribution in redshift
space of a group, $n_{Gg}(r_p,\Delta v)$. The stacked density
distribution is related to the two-dimensional group-galaxy
correlation function, $\xi_{Gg}$ as,
\begin{equation}
n_{Gg}(r_p,\Delta v)=n_Gn_g(z)\left[\xi_{Gg}\left({R\over r_{200}},
{\Delta v\over\sigma_1}\right)+1\right],
\label{eq:xigg} 
\end{equation}
where $n_g(z)$ is a smoothed fit to the redshift distribution locally
converted to velocities in precisely the same way as the real galaxy
pairs. This is the $DD/DR-1$ method \citep{lss} of calculating a
cross-correlation between groups and galaxies. The correlation
function $\xi_{Gg}$ is the fundamental quantity of interest, since
it has the background density removed.  Both the surface density
profile and the velocity dispersion profile are derived from
$\xi_{Gg}$. Note that the total luminosity of the group uses the
magnitude weights, which allows for the declining sampling rate
towards the sample limit. Figure~\ref{fig:xy} builds the confidence
that the selected centers are entirely sensible, although at this
stage we cannot claim that this is an optimal procedure.

To assemble our ``standard'' mean group we will scale all the
velocities to the $\sigma_1$ measured for each group and the radii to
the $r_{200}$ derived from the velocity dispersion. Because there is a
substantial uncertainty in $\sigma_1$ for these small groups there is
a concern that these scalings contain a large random element.  At some
level this is unavoidable. As the galaxies orbit in the groups they
will have different velocities and positions with time which because
of the little averaging available in small groups leads to
considerable variation of their redshift space properties.  Scaling
the velocities and the radii avoids the pitfall of overlaying small,
low velocity dispersion groups on the inside and large, high velocity
dispersion groups on the outside, which would immediately lead to a
rise of velocity dispersion with radius.  The plot of the radial
location of each galaxy that contributes to the mean group as a
function of the source group's velocity dispersion in
Figure~\ref{fig:rr200} does show one potential systematic problem:
the higher velocity dispersion groups do not extend out to
$1.5r_{200}$. This suggests that redshift space interlopers are an
important source of noise leading to velocity dispersions that are too
large for the size of the group. To address this issue we will be
considering a number of alternate groups samples to check our results.

Although these groups have small numbers of members interlopers can be
accurately removed statistically.  We use the smoothed $n_g(z)$ that
we derive from the galaxy sample as a whole using the procedures of
our clustering analysis
\citep{cnoc2_xi}. The random data are scaled in precisely the same way
at the true data to determine whether or not they are in the
cluster. To do the subtraction we need to bin the mean group in $r_p$
and $\Delta v$ bins.  We select bins that are 0.2 dex in
log$_{10}(r_p/r_{200})$ and 0.1 in the velocities scaled to
$\sigma_1$. The binning is needed in our subsequent analysis anyway.
We make velocity distance cuts at 3, 4 and 5 units as a check on the
effect of residual interlopers and normally use the minimal noise 3
unit cut as our standard dataset.

Although we have statistically subtracted the background, the velocity
dispersion estimate is very sensitive to the presence of field
interlopers.  In physical co-ordinates the densities of the groups are
expected to be about $100\rho_0$ near $r_{200}$, but this is reduced
by a factor of roughly $3\sigma_1/[H(z)r_{200}]\simeq 20$ at our
$\Delta v=3\sigma_1$ cutoff. Consequently interlopers, which are
distributed nearly uniformly in distance, generally cause erroneously
large velocity dispersions.  If there is a constant density
normalization underestimate then the fraction of interlopers will
increase with $r_p$, since the contrast between the group and the
surrounding field declines with increasing $r_p$.

The probability of group interlopers can be estimated from the
two-point correlation function.  The probability that a galaxy with
$r_p\le \rpmax$ and $\Delta v<\dvmax$ is physically within $r<r_p$ is
discussed in Carlberg \et\ (2000a). The conclusion is that slightly
more than 50\% of the galaxies within the required velocity separation
are within the required radial separation, $r$. The other 50\% are
clustered towards the group but most probably at roughly a correlation
length.  Other than small factors of order unity this redshift to real
space ambiguity always exists.  This means groups of size $n$ may well
be of size $n-1$. The chances that they are of size $n-2$ drops
precipitously.  Of course as $n$ rises the fraction of interlopers is
falling. These correlated but unvirialized group galaxies cannot be
avoided. Although they present a challenge for a particular group they
are not a significant complication for most statistical analyses of
the mean group, given the mean group density profile we derive below.

\subsection{The Mean Density Profile}

Summing the mean group over velocities gives the mean projected
density profile displayed in Figure~\ref{fig:osurf}. The points are
for cutoffs at 3, 4 and 5 velocity units, with the points declining
slightly at small radii for increasing velocity cutoffs as a result of
the normalization of the total integrated density to unity.  The
errors are evaluated from the patch-to-patch variance. There is little
systematic change with increasing velocity cutoff.  It is immediately
clear that the projected density profile is very nearly a power law
with only a weak break to a slightly shallower central profile. The
mean projected slope of our virialized groups is approximately
$\Sigma_N\propto R^{-1.5}$.  The effect of restricting the sample to
the $\sigma_1 \le 200\kms$ groups is shown in the inset to
Figure~\ref{fig:osurf}. The density profile becomes somewhat steeper
at large radii and there is a suggestion there is a break in the power
law at small radii. The best fit parameters are $a\simeq 0.7$ and
$b\simeq 2.5$, although the core slope is not well defined.

We model the projected galaxy density distribution as the projection
of the galaxy number density distribution,
\begin{equation}
\nu(r) = {A\over {4\pi r^a (r+c)^b}},
\end{equation}
which is projected to a surface density using
\begin{equation}
\Sigma_N(R) = 2\int_R^\infty \nu(r) {r \over\sqrt{r^2-R^2}} \,dr.
\label{eq:surf}
\end{equation}
so that we minimize the variance in the $\Sigma_N(R)$ plane. The resulting
best fit is degenerate because the scale radius is always found to be
very small. For the data displayed we find $c=0.061, 0.074$ and
$0.062$, for cutoffs of 3, 4 and 5, respectively. The $[a,b]$ pairs
are $[0.50,2.05]$, $[0.81,1.71]$, and $[0.78,1.71]$,
respectively. However, other fits with $a+b=2.55\pm 0.05$ are equally
acceptable given the small implied scale radius, $c$.

It is somewhat remarkable that using our mean group center we identify
such a cusped density profile. The procedure is not guaranteed to have
any galaxy at the center of the group, whereas we find that on the
average the galaxy density declines quite steeply away from the
center. This is quite different from the cluster situation where a
similar average center gives a reduced central density
\citep{cnoc1_pro}. The NFW profile does not provide a good fit to
this distribution of galaxies, although it must be borne in mind that
the NFW model describes the dark matter profile, not the galaxy
distribution which can in principle be quite different. Our observed
group profile is near a power law which is steeper in the center and
shallower at large radii, than the NFW function or the steeper central
cusps found by others \citep{moore_core,klypin_gp}.

\subsection{The Mean Velocity Dispersion Profile}

The velocity dispersion profile allows the mass profile to be
constrained. The projected velocity dispersion profile, $\sigma_p(R)$,
for the $\rzmax=5\hmpc$ groups is shown in Figure~\ref{fig:osig}. It
is crucial that the weak rise, or at least absence of of a decline, in
the projected velocity dispersion with radius be real, not an artifact
of inadequate background subtraction. We display the results for
several alternate values of $\rzmax$ in Figure~\ref{fig:osig_rz}, and
find that all give similar velocity dispersion
profiles. Figure~\ref{fig:osig_rz} also shows that setting $\rzmax$
larger than the standard tends to increase the noise, while smaller
values tend to leave out genuine group galaxies in the $\sigma_1$
estimate so that larger velocity cutoffs quickly lead to higher
$\sigma_p(R)$.

To help understand the distribution of possible outliers we present
Figure~\ref{fig:gray}, a gray scale of the $r_p-\Delta v$ plane of the
mean group profile, normalized to $\xi_{Gg}(r_p,\Delta
v)/\Sigma(r_p)$, to remove the surface density variation. The plot
shows that inside of about $0.05r_{200}$ the statistics are too poor
to give useful results.  Beyond about $2r_{200}$ there is no
expectation that the galaxies are in virial equilibrium. The figure
provides evidence that the slow rise in velocity dispersion across the
virialized region is real and not a consequence of increasing
interlopers with radius.  Beyond about $2r_{200}$ virialized motions
are small and the width of the distribution increases rapidly. The
slow rise of $\sigma_p(R)$ is clearly rooted in the data, although a
constant $\sigma_p$ is not strongly excluded.  It is well known that
the pairwise velocity dispersion of galaxies has a similar weak rise
with increasing separation; however that is expected on the basis of
the two-point correlation function falling as $r^{-1.8}$, rather than
the $r^{-2.5}$ that we find here.  A dynamical constraint from the
Jeans equation is that in a scale free distribution, $\sigma^2\propto
\rho(r)r^2$.
Therefore, the surprising outcome is the combination of the ``steep''
$\Sigma(R)$ and ``slowly-rising'' $\sigma_p(R)$.  Interlopers are
always a concern but will tend to increase the outer values of both
these functions, to leave the discrepancy in place.

To evaluate the significance of the radial trends we require reliable
errors estimates. The errors in $\Sigma(R)$ and $\sigma_p(R)$ are
robustly and empirically evaluated from the patch-to-patch variances
in the quantities of interest. We find that the sensitivity to the
velocity cutoff is relatively small, except that the errors increase
in size. As a consequence we will use the 3 velocity unit cutoff as
our standard.  The velocity dispersion is calculated as the second
moment of the binned projected velocity distribution function (pvdf).
Figure~\ref{fig:vdf} displays the average pvdf, normalized to the
velocity dispersion in each radial bin. We note that the observed pvdf
is much closer to Gaussian than to an exponential, although relative
to the Gaussian the group mean pvdf appears to have a small excess at
small and large velocities. This pvdf indicates that only mild
anisotropy is allowed \citep{vdm}, at least for simple distribution
functions.

The projected velocity dispersion is easily modeled as the projection
of a 3D velocity dispersion, where we adjust the radial velocity
dispersion, $\sigma_r(r)$, and the velocity anisotropy, $\beta(r)$, to
fit the data. The projection integral is,
\begin{equation}
\sigma_p^2(R)\Sigma_N(R) = \int_R^\infty \nu(r)\sigma_r^2
        \left(1-\beta(r) {R^2\over r^2}\right) {r\over\sqrt{r^2-R^2}} \,dr.
\label{eq:sigp}
\end{equation}
N-body models suggest that the orbits of galaxies should be similar to
that of the full dark matter distribution, with some deficiency of
radial orbits which are selectively removed due to merging and tidal
destruction \citep{ghigna}.  In principle, the velocity ellipsoid is
constrained by the pvdf, shown in Figure~\ref{fig:vdf}.

We model the radial velocity dispersion as 
\begin{equation}
\sigma_r^2(r)={Br\over b+r},
\label{eq:sigr}
\end{equation}
which is a solution of Jeans equation which worked well for clusters
\citep{cnoc1_pro}. In the absence of other
information, the velocity anisotropy,
$\beta=1-\sigma_\theta^2/\sigma_r^2$, is best taken to be a constant
independent of radius.  However, that choice immediately leads to the
somewhat surprising result that the mass-to-light ratio rises with
radius. In order to investigate the sensitivity of the result to the
velocity anisotropy we adopt the model,
\begin{equation}
\beta(r)=\beta_0 w(r)+\beta_\infty[1-w(r)],
\label{eq:varbeta}
\end{equation}
where $w(r)=r_\beta^2/(r_\beta^2+r^2)$. This $\beta(r)$ function goes
to $\beta_0$ at the origin and $\beta_\infty$ at large radius.  As
reasonable alternatives we set $r_\beta=0.1r_{200}$ or
$0.3r_{200}$. This parameter could in principle be part of the
nonlinear fit, but because dynamical problems are under-constrained
without some knowledge of the velocity anisotropy we choose to use
values of this parameter where our $\beta(r)$ gives a useful variation
over the range of the data.  The fits are shown for the three
different velocity cutoffs in Figure~\ref{fig:osig}.  Note that the
results for different $\beta$ are essentially indistinguishable, and
that the model has little difficulty giving a statistically acceptable
fit to the data. 

\section{The Rising Mass-to-Light Ratio}

The dynamical mass profile can be derived from the tracer density
profile, $\nu(r)$, its velocity dispersion, $\sigma_r(r)$, and the
velocity anisotropy, $\beta(r)$, using Jeans' equation,
\begin{equation}
M(r) = -{\sigma_r^2r\over G}
        \left[{{d \ln{\sigma_r^2}}\over{d\ln{r}}} +
        {{d\ln{\nu}}\over{{d\ln{r}}}} +2\beta\right].
\label{eq:jeans}
\end{equation}
The validity of this equation does not rely on $\nu(r)$ being
distributed like the mass density, $\rho(r)$. It does require that the
system be in dynamical equilibrium, which our group selection
procedures are specifically designed to pick out. Throughout we will
use $L(r) = 4\pi\int \nu r^2 \, dr$ as the luminosity profile,
assuming galaxies have the same mean luminosities at all radii.
Strictly speaking we are working out the mass-to-number ratio. The
assumption that low and high luminosity galaxies are similarly
distributed relative to the mass field is explicitly tested below. The
advantage of this approach is that it weights the galaxies relatively
equally, rather than concentrating most of the statistical weight on
the high luminosity galaxies.

In Figure~\ref{fig:mol} we display the $M(r)/L(r)$ profile for the
standard group centers with the velocity dispersion calculation cutoff
set at three velocity units.  Note that this shows the integrated
interior $M/L$, not the local values.  We first model these standard
groups using a conventional approach with minimal anisotropy. That is,
the central velocity ellipsoid is isotropic, \ie\ $\beta_0=0$, the
anisotropy radius, $r_\beta$, is at about the midpoint of the
virialized region, $r_\beta=0.3r_{200}$. The outer anisotropy,
$\beta_\infty$, takes on the values $-1,-\onehalf,0$ and
$\onequarter$, which range from nearly tangential to slightly radial
velocity anisotropy.  The $M/L$ curves are normalized to unity at the
data point closest to $r/r_{200}=1$. The data indicate about a factor
of three to ten rise in the mass-to-light per decade of radius, with
little sensitivity to the outer velocity anisotropy. That is, the
light is much more centrally concentrated than the mass. This effect
is visible for many of the groups where their sky distribution is much
more concentrated than their $r_{200}$ as shown in
Figure~\ref{fig:xy}.

The insets in Figure~\ref{fig:mol} address the concern about
erroneously large $\sigma_1$ values biasing the properties of the mean
groups. The lower inset shows the results of the analysis based on the
stacked groups restricted to have $\sigma_1 \le 200 \kms$. The upper
inset is the result of the analysis where we have overlaid all groups
in physical co-ordinates.  For convenience and comparability in this
case we have taken $\sigma_1=200 \kms$ for the scaling. Although there
are differences in the details of the result $M/L$ curves the rising
trend is preserved. Overlaying groups in physical space serves to
minimize the size of the rise. The differences between these results
serves as an indication of the size of the systematic errors.

The mass-density profile, $\rho(r)$, derived from the mass profile as
$(4\pi r^2)^{-1} dM/dr$, is plotted in Figure~\ref{fig:den} for the
same models as in Figure~\ref{fig:mol}.  The model with the most
extreme velocity ellipsoid at large radii, the tangentially dominant
$\beta_\infty=-1$ (solid line), implies a large core which becomes
relatively smaller as $\beta_\infty$ becomes more positive. This model
dependent result is one of the most interesting aspects of these
groups and its reality needs to be tested which we can at least partially
do within our own data. The upper and lower insets are for the
$\sigma_1\le 200 \kms$ sample and the physical co-ordinate stacking of
the entire sample, respectively.

An alternate version of the groups, with $\rzmax=3\hmpc$, is analyzed
in the same way as the $\rzmax=5\hmpc$ groups with mass-to-light shown
in the upper left panel of Figure~\ref{fig:omol} and density in
Figure~\ref{fig:oden}.  In this case, an even steeper rise of
mass-to-light emerges than for the standard groups. For velocity
ellipsoids with $\beta_\infty>-1$ the central mass density is negative.
The effect of analyzing the standard groups with $r_\beta=0.1r_{200}$
is shown in the upper right panels of Figures~\ref{fig:omol} and
\ref{fig:oden}. This small $r_\beta$ boosts the sensitivity of
the gradient to $\beta_\infty$.  We conclude that the existence of a
core in the dark matter distribution depends sensitively on the
assumption that the velocity anisotropy is close to isotropic.

A strong tangential anisotropy in the center, $\beta_0=-1$, is able to
almost eliminate the mass-to-light gradient as shown in the lower
panels of Fig~\ref{fig:oden}, with $r_\beta=0.3r_{200}$ (left) and
$r_\beta=0.1r_{200}$ (right).  However we recall that the pvdf of
Figure~\ref{fig:vdf} does not favor such strong anisotropy. The
implied density, shown in the lower left panel of Fig~\ref{fig:oden},
is nearly a power law, $r^{-2}$, for $\beta_\infty=-1$, and none of
these resemble the predicted dark matter density profiles
\citep{nfw,moore_core}.

The source of the rising $M/L$ in the kinematic data is readily
understood. In a power law density distribution with a radial density
dependence of $\rho\propto r^{-2+p}$, the mass generating the potential
must have a velocity dispersion profile of $\sigma^2 \propto r^p$.  From
the velocity dispersion we measure that $p\simeq
\onequarter\pm\onequarter$. This implies a mass density profile that
is slightly shallower than $r^{-2}$. The measured mean light profile,
$\nu(r)\propto r^{-2+q}$, with $q\simeq -\onehalf$. Consequently the
mass to light ratio, $\int \rho r^2\, dr/\int \nu r^2\,dr$, which
varies as $r^{p-q}$, rises as $r^{\threequarters\pm\onequarter}$.
Nearly circular orbits can flatten the $M/L$ profile for the observed
velocities, but these $\beta$ would be very different than the
observational results for rich galaxy clusters
\citep{cnoc1_pro,vdm} and n-body experiments
\citep{ghigna}.  One significant difference between groups and rich clusters
is that the timescale for dynamical friction to act declines as
$\sigma_1^3$, so that dynamical friction is much more effective for
galaxies orbiting in groups than clusters.

\subsection{Color and Luminosity Dependent $M/L$ Profiles}

As a test of the galaxy population dependence of the result, we divide
the group galaxies in to blue and a red subsamples, splitting at
$(B-R)_0=1.25$ mag. In a second test we divide the group galaxy sample
into high and low luminosity subsamples, splitting at $M_R^{k,e}=-20$
mag.  The resulting $M/L$ profiles, both masses and luminosity
profiles calculated from these subsamples alone, are shown in
Figure~\ref{fig:omol_sub}. To simplify the comparison we keep
$\beta_0=0$ for all models, but continue to vary $\beta_\infty$.

The blue galaxies do a much better job of tracing the mass profile,
that is, it is fairly flat, than the red galaxies do.  The opposite is true
in clusters \citep{cnoc1_pro}.  In both clusters and groups the red
galaxies are more centrally concentrated than the blue galaxies.  The
$M/L$ gradient does not have a significant dependence on the
luminosities of the galaxies. The latter is an argument that dynamical
friction on a galaxy's gas and stars might not be the dominant source
of the segregation of mass and light in groups.  However, the presence
of a much stronger $M/L$ gradient for the red galaxies would be
consistent with dynamical friction taking several orbits after entry
into the cluster to produce a segregation.

\section{Discussion and Conclusions}

In this paper we use the CNOC2 survey catalogues, containing
approximately 6000 galaxies with redshifts, to identify more than 200
high probability virialized groups containing, on the average, 3.8
members with redshifts for a total of about 1000 group members. This
represents about 25\% of all eligible galaxies in the sample. These
groups have selection dependent $M_{VT}/L$ values comparable to the
values of large clusters with the ``standard'' groups being about a
factor of two lower. The group population volume density matches on to
the number-velocity dispersion relation of clusters. The clustering of
groups is enhanced at about the level expected for the approximately
200 \kms\ velocity dispersion dark matter halos of groups as compared
to the approximately 100 \kms\ halos of individual galaxies.  Overall
the global properties of groups are about as expected for dark matter
halos on this mass scale.

Quite unlike rich clusters which have a mass-to-light profile constant
with radius \citep{cnoc1_pro}, our analysis of the internal properties
of galaxy groups finds considerable evidence for a gently rising
mass-to-light profile with radius. A second difference is that in
groups the blue galaxies follow the dark mass more closely than the
red galaxies. This paper presents several lines of argument that the
rising velocity dispersion profile with radius is not an artifact of
the analysis. Moreover, the projected velocity distribution function
indicates that the orbital distribution is unlikely to consist almost
entirely of circular orbits, as would be required to minimize the rise
in $M/L$ with radius.

The inference that $M/L$ rises with radius depends sensitively on the
theoretical and indirect observational indication that the 
mean velocity ellipsoid is approximately isotropic.  Hence this result
needs to be independently verified.  Current indirect support for this
result comes from X-ray observations of low redshift galaxy groups
having velocity dispersion overlap with our sample near 300
\kms.  The X-ray emission is 
significantly more extended than most of the group light indicating a
rising $M/L$ \citep{kcc,mz}.  Weak gravitational lensing is one of the
best prospects to check this result since it has no dependence on
assumptions of equilibrium. These intermediate redshift groups are
ideally situated for weak lensing studies \citep{henk}.  The somewhat
puzzling weak lensing results for the low velocity dispersion cluster
MS1224+20 in two independent studies indicates that the $M/L$ is
rising with radius \citep{fksw,vb_1224,fischer}. Given the large group
to group fluctuations it will be important to examine a statistically
meaningful sample.

A highly model dependent result is that we find that the mean group
most likely has a significant core radius.  To retain a power law form
would require very tangential orbits near the center. However, a large
core could simply be a trivial result of the group galaxies sloshing
around in the group potential, such that the centers of light and mass
are not coincident. To further investigate the location and radial
dependence of the central density profile of groups will require
studying individual groups in detail, possibly with X-ray and weak
lensing measurements.

What is the origin of the rising $M/L$? The two physical possibilities
are that galaxies sink with respect to the dark matter, or, that group
dark matter has a small effective pressure which does not allow the
core to undergo the same gravitational collapse as cold collisionless matter
can. It is important to note that near $r_{200}$ both the galaxies and
the dark matter have undergone about the same amount of collapse
relative to the field. It is only in the inner third or so where the
major differences develop, and the size of those differences is
strongly dependent on the kinematic model.  The simplest explanation
is that these observations confirm the standard picture of a
collisionless dark matter with gaseous dissipation allowing the
baryonic matter to concentrate to the center of individual dark matter
halos.  As galaxies cluster together to form a group their individual
halos are tidally removed to join the common halo from the outside in.
Dynamical friction from the collisionless dark matter causes the
galaxies to sink in the common group halo, over about a Hubble time
\cite{barnes,mamon,evrard,bcl,peb}.  One possible concern with this
interpretation is that when groups join together to make rich clusters
this history needs to be erased to leave no measurable $M/L$ gradient.
Furthermore, whether this mechanism is consistent with no luminosity
dependence of the $M/L$ gradient but a large galaxy color dependence
of the $M/L$ gradient puts some fairly strong, but not necessarily
unreasonable constraints, on the formation history of the galaxies.

A more extreme possibility for the origin of the $M/L$ gradient is
that the dark matter is subject to some effective pressure that does
not allow it to undergo full gravitational collapse to form a core,
either through a phase density limit, or through collisional interactions
\citep{ss}. More conventional explanations should be carefully
examined and our results independently verified with other methods of
observation before this is accepted.

\acknowledgments

This research was supported by NSERC and NRC of Canada.  HL 
acknowledges support provided by NASA through Hubble Fellowship grant
\#HF-01110.01-98A awarded by the Space Telescope Science Institute,
which is operated by the Association of Universities for Research in
Astronomy, Inc., for NASA under contract NAS 5-26555.  We thank the
CFHT Corporation for support, and the operators for their enthusiastic
and efficient control of the telescope.

~
Table 1: The $\rzmax=5\hmpc$, $\rpmax=0.25\hmpc$ Groups
\scriptsize

\begin{tabular}{|r|r|r|l|l|r|l|}\hline
x & y & z & $\sigma_1$ & $M/L$ &  $N_g^z$ & $M$  \\ 
$\arcsec$ &$\arcsec$ & & $\kms$ & $h\msun/\lsun$& & $h^{-1}\msun$ \\
	\hline
\hline
\multicolumn{7}{|c|}{0223+00} \\ \hline\hline
  -256.4&   546.7&  0.18809&     80$\pm$    70&  11$\pm$  36&   3& 5.116e+11$\pm$ 192\% \\
  -204.3&  1484.6&  0.19834&     95$\pm$    74&  58$\pm$ 324&   3& 2.303e+12$\pm$ 127\% \\
  1589.7& -1504.7&  0.21787&    126$\pm$    71&  41$\pm$  61&   5& 3.443e+12$\pm$ 100\% \\
   -16.2&  2522.1&  0.22075&     65$\pm$    68&  33$\pm$ 106&   3& 1.349e+12$\pm$ 183\% \\
   549.8&  -171.4&  0.22781&    111$\pm$    56&  39$\pm$  56&   4& 3.995e+12$\pm$  73\% \\
   382.6&  -110.8&  0.22920&    184$\pm$    50& 206$\pm$ 190&   4& 8.637e+12$\pm$  48\% \\
   117.7&  -395.2&  0.26741&    254$\pm$   104& 274$\pm$ 283&   6& 2.084e+13$\pm$  76\% \\
  2367.5& -1396.8&  0.27011&    134$\pm$    38& 108$\pm$ 104&   6& 5.681e+12$\pm$  74\% \\
   162.9&  1737.6&  0.27038&    148$\pm$    86& 262$\pm$ 433&   4& 7.775e+12$\pm$ 101\% \\
  2331.2& -1613.0&  0.27045&    118$\pm$   123& 104$\pm$ 214&   3& 4.478e+12$\pm$ 137\% \\
   218.8&   708.6&  0.29864&    115$\pm$    66&  30$\pm$  44&   5& 1.704e+12$\pm$  94\% \\
  -753.8&  -146.8&  0.30207&    375$\pm$    85& 271$\pm$ 339&   5& 3.367e+13$\pm$  43\% \\
   -44.3&  -884.3&  0.30275&     95$\pm$   103&  62$\pm$ 108&   3& 1.541e+12$\pm$ 123\% \\
  -130.1&  2991.5&  0.30496&    135$\pm$   130&  51$\pm$  68&   3& 3.504e+12$\pm$ 115\% \\
   132.8&  2579.2&  0.30520&    266$\pm$   181& 143$\pm$ 202&   5& 1.619e+13$\pm$ 109\% \\
   256.6&    74.7&  0.30647&    305$\pm$   326&1115$\pm$2313&   3& 3.278e+13$\pm$ 150\% \\
   554.8&    64.3&  0.30933&     95$\pm$    73&  49$\pm$  74&   4& 2.304e+12$\pm$ 116\% \\
  1198.3& -1408.6&  0.33835&     92$\pm$    53&  13$\pm$  25&   4& 1.092e+12$\pm$ 165\% \\
  -105.8&  1465.8&  0.35084&    160$\pm$   197& 132$\pm$ 406&   3& 5.871e+12$\pm$ 179\% \\
   -41.2&  -882.2&  0.35746&    122$\pm$   112&  46$\pm$  92&   3& 4.951e+12$\pm$ 113\% \\
   -92.1&  -736.2&  0.35772&    154$\pm$   150& 140$\pm$ 279&   4& 1.130e+13$\pm$ 131\% \\
   450.5&  -386.1&  0.35811&    574$\pm$   483&1557$\pm$2715&   4& 1.102e+14$\pm$ 117\% \\
   710.7&  -451.5&  0.35851&    305$\pm$   104& 120$\pm$ 139&   7& 2.441e+13$\pm$  74\% \\
   608.7&   -87.3&  0.36049&    365$\pm$   169& 693$\pm$1044&   5& 6.553e+13$\pm$  80\% \\
   557.6&  -226.1&  0.36107&    202$\pm$    68&  85$\pm$ 155&   5& 1.190e+13$\pm$  62\% \\
  -396.9&  -324.0&  0.36464&    265$\pm$   203& 289$\pm$ 498&   3& 3.008e+13$\pm$  92\% \\
   929.9& -1595.8&  0.38473&    362$\pm$   218& 287$\pm$ 371&   5& 4.703e+13$\pm$  88\% \\
  -642.4&  -609.8&  0.38483&    180$\pm$   193& 131$\pm$ 345&   3& 8.168e+12$\pm$ 196\% \\
  1016.4&  -856.2&  0.38644&    228$\pm$   173& 119$\pm$ 155&   6& 1.500e+13$\pm$ 104\% \\
   631.4& -1119.5&  0.38661&    229$\pm$   165& 107$\pm$ 172&   4& 1.064e+13$\pm$ 128\% \\
   102.1&  2003.8&  0.39651&    216$\pm$    42& 127$\pm$  84&   6& 1.780e+13$\pm$  35\% \\
  1639.8& -1625.9&  0.39654&    645$\pm$   443&2439$\pm$6437&   3& 1.521e+14$\pm$ 191\% \\
  -102.1&   609.7&  0.39715&    261$\pm$   211& 264$\pm$ 484&   3& 2.087e+13$\pm$ 103\% \\
  -167.6&  2053.8&  0.39785&    230$\pm$   174& 207$\pm$ 170&   4& 2.036e+13$\pm$  76\% \\
    89.5&  1582.7&  0.40166&    191$\pm$    92& 128$\pm$ 164&   4& 1.475e+13$\pm$  66\% \\
   233.2&  1396.3&  0.40494&    667$\pm$   479&3266$\pm$14005&   3& 1.799e+14$\pm$ 252\% \\
   695.3& -1015.4&  0.40785&    459$\pm$   141& 277$\pm$ 370&   5& 7.083e+13$\pm$  54\% \\
\hline
\end{tabular}\newpage
\begin{tabular}{|r|r|r|l|l|r|l|}\hline
x & y & z & $\sigma_1$ & $M/L$ &  $N_g^z$ & $M$  \\ 
$\arcsec$ &$\arcsec$ & & $\kms$ & $h\msun/\lsun$& & $h^{-1}\msun$ \\
	\hline
\hline
\multicolumn{7}{|c|}{0223+00} \\ \hline\hline
   566.7&  -232.9&  0.40816&    444$\pm$   522& 551$\pm$1976&   3& 6.454e+13$\pm$ 231\% \\
   464.6&   197.9&  0.41259&    234$\pm$   247& 134$\pm$ 196&   3& 1.003e+13$\pm$ 126\% \\
    56.1&  2154.8&  0.41916&    183$\pm$    60& 140$\pm$ 124&   5& 1.484e+13$\pm$  50\% \\
  -171.5&  2316.5&  0.44237&     86$\pm$    94&  46$\pm$ 132&   3& 2.524e+12$\pm$ 171\% \\
    29.0&  3007.5&  0.46898&    197$\pm$   142&  64$\pm$ 192&   3& 1.091e+13$\pm$  92\% \\
   177.0&  2967.2&  0.47066&     79$\pm$    58&   8$\pm$  28&   3& 7.906e+11$\pm$ 177\% \\
\hline
\multicolumn{7}{|c|}{0920+37}\\ \hline\hline
  -243.5&    14.5&  0.19117&    253$\pm$   129& 215$\pm$ 550&   5& 1.928e+13$\pm$ 104\% \\
   182.0&  1070.6&  0.19149&     99$\pm$    82&  82$\pm$ 230&   3& 3.290e+12$\pm$ 109\% \\
   130.7&  -604.8&  0.20203&    108$\pm$   104&  67$\pm$ 299&   3& 3.422e+12$\pm$ 284\% \\
  -526.3& -1112.5&  0.20227&    143$\pm$   163& 124$\pm$ 277&   3& 3.418e+12$\pm$ 146\% \\
     0.5&  1098.7&  0.20714&    262$\pm$   170& 150$\pm$ 343&   4& 1.120e+13$\pm$ 108\% \\
  2342.2& -1435.5&  0.22135&    392$\pm$   120& 299$\pm$ 375&   8& 4.539e+13$\pm$  70\% \\
  2294.1& -1439.0&  0.22476&    164$\pm$    96& 168$\pm$ 302&   4& 1.029e+13$\pm$  96\% \\
  2406.8& -1450.4&  0.22506&    182$\pm$   188& 342$\pm$1308&   3& 1.077e+13$\pm$ 130\% \\
  2339.5& -1290.9&  0.22556&    213$\pm$   255& 509$\pm$1235&   3& 1.584e+13$\pm$ 170\% \\
  -172.0&    89.5&  0.23083&     83$\pm$    54&  24$\pm$  40&   4& 2.096e+12$\pm$ 114\% \\
    31.2&  -266.9&  0.23299&    197$\pm$   146& 306$\pm$ 889&   3& 1.046e+13$\pm$  96\% \\
  -200.0&  -673.6&  0.24328&    157$\pm$   104& 223$\pm$ 423&   4& 7.858e+12$\pm$ 105\% \\
  -325.3&  -274.7&  0.24347&    202$\pm$   107& 598$\pm$ 947&   4& 1.665e+13$\pm$  90\% \\
  -166.2&   189.2&  0.24395&    148$\pm$    43&  90$\pm$ 132&   6& 7.245e+12$\pm$  59\% \\
   -17.3&   931.5&  0.24438&    185$\pm$    49& 235$\pm$ 192&   6& 1.242e+13$\pm$  56\% \\
   303.9&   103.3&  0.24441&    158$\pm$    27&  81$\pm$  52&   7& 6.221e+12$\pm$  42\% \\
   -40.4&  -388.1&  0.24447&    125$\pm$   105& 210$\pm$ 866&   3& 4.561e+12$\pm$ 256\% \\
  -258.2&   308.6&  0.24500&    348$\pm$   329&1209$\pm$2431&   3& 3.154e+13$\pm$ 118\% \\
   151.1&  1534.7&  0.24515&    297$\pm$   195& 482$\pm$1684&   4& 2.833e+13$\pm$  94\% \\
  -126.4&  2379.0&  0.24535&    226$\pm$    97& 145$\pm$ 157&   5& 1.272e+13$\pm$  79\% \\
    91.4&  1894.1&  0.24611&    263$\pm$    58& 337$\pm$ 351&   5& 2.289e+13$\pm$  46\% \\
  -607.9&   178.4&  0.24623&    147$\pm$    47&  58$\pm$  83&   6& 4.670e+12$\pm$  73\% \\
   178.9&   280.5&  0.24763&    190$\pm$   107& 199$\pm$ 406&   4& 1.366e+13$\pm$ 119\% \\
  1315.8&  -902.3&  0.25424&     97$\pm$    78&  35$\pm$ 180&   3& 2.415e+12$\pm$ 378\% \\
    36.6&   749.5&  0.25955&    192$\pm$   139& 114$\pm$ 176&   6& 1.369e+13$\pm$ 119\% \\
   983.1& -1531.8&  0.28641&     72$\pm$    60&  25$\pm$  69&   3& 4.023e+11$\pm$ 169\% \\
    61.6&   693.3&  0.31835&    427$\pm$   473&3073$\pm$8122&   3& 9.172e+13$\pm$ 152\% \\
   381.8&  -149.0&  0.32181&    144$\pm$   176& 205$\pm$ 649&   3& 5.857e+12$\pm$ 181\% \\
  -145.8&  1587.5&  0.32201&    111$\pm$    93&  33$\pm$ 117&   3& 2.974e+12$\pm$ 113\% \\
    74.4&  1020.4&  0.32273&    167$\pm$   149& 136$\pm$ 367&   4& 9.154e+12$\pm$ 170\% \\
\hline
\end{tabular}

\newpage
\begin{tabular}{|r|r|r|l|l|r|l|}\hline
x & y & z & $\sigma_1$ & $M/L$ &  $N_g^z$ & $M$  \\ 
$\arcsec$ &$\arcsec$ & & $\kms$ & $h\msun/\lsun$& & $h^{-1}\msun$ \\
	\hline
\multicolumn{7}{|c|}{0920+37}\\ \hline\hline
   203.4&   403.9&  0.32380&    230$\pm$   113& 305$\pm$ 405&   4& 1.378e+13$\pm$  98\% \\
  -169.5&  1189.9&  0.32384&    256$\pm$   128& 472$\pm$ 543&   5& 2.173e+13$\pm$  92\% \\
   -20.4&   721.8&  0.36173&    109$\pm$    92&  66$\pm$ 290&   3& 4.329e+12$\pm$ 312\% \\
   184.0&  -816.1&  0.36178&    412$\pm$   507& 596$\pm$1514&   3& 3.697e+13$\pm$ 181\% \\
   114.1&  2543.7&  0.37218&    542$\pm$   308& 838$\pm$1372&   4& 5.252e+13$\pm$ 117\% \\
  -175.3&   180.4&  0.37237&    160$\pm$    32&  68$\pm$  79&   4& 4.896e+12$\pm$  54\% \\
  -155.8& -1194.7&  0.37247&    436$\pm$   148& 502$\pm$ 427&   6& 5.015e+13$\pm$  56\% \\
  -233.7&   -90.2&  0.37291&    126$\pm$   125&  75$\pm$ 128&   3& 6.057e+12$\pm$ 123\% \\
  -370.3&  -888.2&  0.37317&    328$\pm$   210& 360$\pm$ 607&   5& 4.321e+13$\pm$  95\% \\
  -119.8& -1090.8&  0.37323&    251$\pm$    85& 284$\pm$ 425&   4& 1.561e+13$\pm$  90\% \\
  -376.8&  -132.5&  0.37331&    231$\pm$   129& 287$\pm$ 832&   4& 1.720e+13$\pm$ 197\% \\
  -354.6&  -742.9&  0.37352&    147$\pm$    81&  60$\pm$  81&   4& 1.194e+13$\pm$ 127\% \\
  -255.8&  -962.7&  0.37406&     86$\pm$    82&  43$\pm$  75&   3& 2.761e+12$\pm$ 113\% \\
  -218.3& -1056.9&  0.37426&    639$\pm$   154& 761$\pm$ 628&   8& 1.587e+14$\pm$  49\% \\
  -102.3&   226.3&  0.37597&    541$\pm$   506&1085$\pm$2781&   3& 6.384e+13$\pm$ 125\% \\
  1759.2& -1385.0&  0.37893&    109$\pm$   114&  78$\pm$ 240&   3& 5.958e+12$\pm$ 120\% \\
   742.9&  -481.4&  0.37894&    353$\pm$   246& 402$\pm$1069&   3& 2.454e+13$\pm$ 112\% \\
  1036.3&  -971.8&  0.37903&    281$\pm$   224& 298$\pm$1408&   3& 3.073e+13$\pm$ 113\% \\
   -54.4&   661.1&  0.37978&     91$\pm$   102&  39$\pm$ 111&   3& 2.769e+12$\pm$ 121\% \\
  -366.6&  -493.7&  0.38032&    328$\pm$   377& 331$\pm$ 677&   3& 1.638e+13$\pm$ 135\% \\
  -257.3&  1320.1&  0.38462&    138$\pm$   104&  82$\pm$ 206&   3& 9.262e+12$\pm$ 177\% \\
   383.8&    91.6&  0.38820&    403$\pm$   108& 492$\pm$ 542&   4& 5.753e+13$\pm$  55\% \\
   900.0& -1302.6&  0.38995&    232$\pm$   191& 510$\pm$1099&   3& 3.083e+13$\pm$ 174\% \\
    86.5&   426.7&  0.39019&    158$\pm$   122&  56$\pm$ 178&   3& 4.814e+12$\pm$ 112\% \\
  1720.0& -1718.3&  0.39096&    184$\pm$   195& 103$\pm$ 192&   4& 1.537e+13$\pm$ 120\% \\
   445.1&   201.0&  0.39177&     86$\pm$    54&  18$\pm$  34&   4& 2.451e+12$\pm$ 109\% \\
   114.2&   112.1&  0.39196&    209$\pm$   166& 210$\pm$ 573&   3& 1.562e+13$\pm$ 100\% \\
  -381.9&   183.6&  0.39527&    181$\pm$   199& 142$\pm$ 200&   3& 5.827e+12$\pm$ 122\% \\
   569.1&  -911.0&  0.42748&    102$\pm$    88&  48$\pm$  76&   3& 3.825e+12$\pm$ 117\% \\
   974.2& -1696.5&  0.46076&    445$\pm$   484& 270$\pm$1094&   3& 3.673e+13$\pm$ 125\% \\
  -308.3&   183.0&  0.46254&    324$\pm$   379& 200$\pm$ 514&   3& 3.258e+13$\pm$ 171\% \\
   -32.5&  -127.0&  0.47277&    294$\pm$   212& 201$\pm$ 430&   3& 2.731e+13$\pm$ 156\% \\
  -568.7&  -193.4&  0.47289&    108$\pm$   100&  26$\pm$  58&   3& 2.635e+12$\pm$ 114\% \\
   971.2& -1332.8&  0.47318&    353$\pm$   204& 139$\pm$ 223&   4& 3.848e+13$\pm$  93\% \\
\hline
\multicolumn{7}{|c|}{1447+09} \\ \hline\hline
   -62.1&  -334.7&  0.16531&    164$\pm$   126& 322$\pm$ 462&   3& 1.263e+13$\pm$ 126\% \\
   -28.5&   840.0&  0.19733&    233$\pm$   170& 318$\pm$ 543&   4& 1.466e+13$\pm$ 136\% \\
\hline
\end{tabular}

\newpage
\begin{tabular}{|r|r|r|l|l|r|l|}\hline
x & y & z & $\sigma_1$ & $M/L$ &  $N_g^z$ & $M$  \\ 
$\arcsec$ &$\arcsec$ & & $\kms$ & $h\msun/\lsun$& & $h^{-1}\msun$ \\
	\hline
\multicolumn{7}{|c|}{1447+09} \\ \hline\hline
  -134.2&  1877.9&  0.20204&    326$\pm$   360&1328$\pm$3276&   3& 2.271e+13$\pm$ 151\% \\
   155.4&  1998.1&  0.20249&    250$\pm$    87& 190$\pm$ 438&   4& 1.066e+13$\pm$ 168\% \\
  -207.0&  2461.4&  0.22454&    256$\pm$   280& 248$\pm$1706&   3& 1.527e+13$\pm$ 140\% \\
    12.6&  1767.1&  0.22854&    268$\pm$   107& 811$\pm$1911&   4& 3.147e+13$\pm$ 102\% \\
  -120.9&  1254.9&  0.22893&     81$\pm$    98&  37$\pm$  93&   3& 2.308e+12$\pm$ 165\% \\
   629.1&    26.7&  0.22897&    162$\pm$   139& 325$\pm$1186&   3& 7.214e+12$\pm$ 113\% \\
   595.9&  -676.4&  0.26162&    229$\pm$    77& 239$\pm$ 668&   4& 1.156e+13$\pm$  49\% \\
   142.6&  -986.1&  0.26981&    104$\pm$    92& 119$\pm$ 378&   3& 2.983e+12$\pm$ 132\% \\
   361.6& -1090.1&  0.27057&    112$\pm$    66&  68$\pm$ 132&   4& 3.020e+12$\pm$ 118\% \\
   522.3& -1144.8&  0.27177&    280$\pm$   143& 347$\pm$ 399&   6& 1.969e+13$\pm$  85\% \\
   -90.6&  -641.5&  0.27286&    165$\pm$   120& 241$\pm$ 732&   3& 8.104e+12$\pm$ 110\% \\
  -693.8&   189.9&  0.28246&    231$\pm$   116& 187$\pm$ 318&   5& 1.367e+13$\pm$  80\% \\
   728.2&  -632.0&  0.30639&     93$\pm$    65&  34$\pm$ 126&   3& 1.329e+12$\pm$ 309\% \\
  -622.3&  -135.7&  0.30643&    199$\pm$   161& 228$\pm$ 653&   4& 1.598e+13$\pm$ 103\% \\
  1091.1& -1288.5&  0.31031&     83$\pm$    85&   6$\pm$  29&   3& 6.854e+11$\pm$ 443\% \\
  1257.0&  -775.6&  0.32378&    180$\pm$   149& 181$\pm$ 375&   4& 8.494e+12$\pm$ 145\% \\
  -526.0&  -639.8&  0.32506&    175$\pm$   151& 125$\pm$ 214&   4& 9.875e+12$\pm$ 114\% \\
  -241.0&   655.4&  0.32716&    247$\pm$   148& 144$\pm$ 232&   5& 1.542e+13$\pm$  96\% \\
  -216.5&  1827.7&  0.34855&    136$\pm$    70&  50$\pm$  76&   6& 4.741e+12$\pm$ 113\% \\
   -98.0&  1364.0&  0.35043&    168$\pm$   169& 200$\pm$ 377&   3& 8.209e+12$\pm$ 120\% \\
   158.1&  1350.7&  0.35062&    250$\pm$   243& 147$\pm$ 239&   3& 1.893e+13$\pm$ 126\% \\
   517.5&   -83.8&  0.35914&     82$\pm$    66&  16$\pm$  27&   4& 1.646e+12$\pm$ 127\% \\
   -80.2&   306.3&  0.36192&    170$\pm$    95& 131$\pm$ 183&   5& 9.828e+12$\pm$ 120\% \\
  -432.0&   197.2&  0.36404&     77$\pm$    94&  41$\pm$ 108&   3& 1.946e+12$\pm$ 173\% \\
   -38.7&  -364.1&  0.37225&    126$\pm$    94&  69$\pm$ 153&   3& 2.787e+12$\pm$ 131\% \\
  -661.0&  -422.0&  0.37263&     44$\pm$    41&   4$\pm$   9&   4& 2.115e+11$\pm$ 154\% \\
  -166.4& -1015.4&  0.37389&    291$\pm$   193& 792$\pm$1798&   4& 3.386e+13$\pm$ 158\% \\
  -338.2& -1151.5&  0.39369&    308$\pm$   257& 438$\pm$ 535&   4& 3.596e+13$\pm$ 110\% \\
   359.7&   174.4&  0.39369&    394$\pm$   406& 864$\pm$1972&   3& 4.819e+13$\pm$ 137\% \\
  -256.3&  -726.8&  0.39438&    507$\pm$   469&1254$\pm$2636&   3& 7.807e+13$\pm$ 114\% \\
    51.3&  1236.2&  0.40653&    101$\pm$    86&  54$\pm$  76&   3& 3.620e+12$\pm$ 103\% \\
   836.6& -1022.5&  0.46575&    150$\pm$   146& 118$\pm$ 215&   3& 6.751e+12$\pm$ 122\% \\
    81.2&  -278.7&  0.46820&    488$\pm$   417& 668$\pm$1154&   4& 7.482e+13$\pm$ 125\% \\
   124.3&    48.3&  0.46930&    217$\pm$   224&  78$\pm$ 166&   4& 8.894e+12$\pm$ 141\% \\
   101.0&  -194.4&  0.47168&    123$\pm$    99&  40$\pm$  70&   3& 4.871e+12$\pm$ 121\% \\
   217.5&  -572.5&  0.51077&    565$\pm$   668& 767$\pm$1710&   3& 1.237e+14$\pm$ 149\% \\
   169.5&  1282.7&  0.53680&    520$\pm$   203& 477$\pm$ 625&   5& 1.355e+14$\pm$  72\% \\
   213.1&  -782.7&  0.54249&    151$\pm$   132&  63$\pm$ 197&   3& 8.662e+12$\pm$ 177\% \\
\hline
\end{tabular}\newpage
\begin{tabular}{|r|r|r|l|l|r|l|}\hline
x & y & z & $\sigma_1$ & $M/L$ &  $N_g^z$ & $M$  \\ 
$\arcsec$ &$\arcsec$ & & $\kms$ & $h\msun/\lsun$& & $h^{-1}\msun$ \\
	\hline
\multicolumn{7}{|c|}{2148-05} \\ \hline\hline
   116.4&  2726.2&  0.15480&    257$\pm$    76& 211$\pm$ 598&   6& 2.419e+13$\pm$  71\% \\
   -14.8&  2646.0&  0.15649&    241$\pm$   129& 164$\pm$ 298&   4& 2.631e+13$\pm$ 102\% \\
  -181.0&  -992.4&  0.17776&    173$\pm$   193& 167$\pm$ 490&   3& 6.127e+12$\pm$ 124\% \\
    32.9&  -660.2&  0.17803&    238$\pm$   182& 767$\pm$1765&   3& 3.069e+13$\pm$ 100\% \\
   119.0&   949.8&  0.19828&    267$\pm$   313& 356$\pm$1741&   3& 1.482e+13$\pm$ 439\% \\
  -181.8&  3095.2&  0.21406&    152$\pm$   111& 306$\pm$ 611&   3& 1.124e+13$\pm$ 111\% \\
  -377.8&    11.8&  0.21926&    275$\pm$    81& 273$\pm$ 317&   7& 2.267e+13$\pm$  65\% \\
  -258.5&  -163.1&  0.21940&    192$\pm$    69&  96$\pm$  80&   6& 5.039e+12$\pm$  71\% \\
  -235.8&   151.0&  0.21941&    160$\pm$    31&  34$\pm$  51&   7& 7.099e+12$\pm$  52\% \\
  -486.0&  -218.9&  0.21964&     96$\pm$    45&  30$\pm$  75&   4& 1.494e+12$\pm$ 111\% \\
   452.9& -1101.2&  0.23645&    121$\pm$    90&  73$\pm$ 224&   3& 3.969e+12$\pm$ 218\% \\
  -765.5&   -98.4&  0.24135&    149$\pm$    38&  46$\pm$  59&   5& 3.683e+12$\pm$  41\% \\
  -179.3&  1552.7&  0.24322&    274$\pm$   168& 316$\pm$1951&   4& 3.163e+13$\pm$ 108\% \\
  -134.4&  2245.3&  0.24344&    146$\pm$   114& 113$\pm$ 174&   3& 7.044e+12$\pm$ 124\% \\
   466.5&  -833.0&  0.26489&    306$\pm$   176& 261$\pm$ 334&   6& 3.355e+13$\pm$  90\% \\
   206.1&  2896.4&  0.28672&    150$\pm$    74& 111$\pm$ 152&   4& 9.487e+12$\pm$  65\% \\
    57.6&  2932.0&  0.28792&    144$\pm$   114& 217$\pm$ 751&   3& 9.609e+12$\pm$ 201\% \\
  -783.8& -1131.2&  0.30139&     70$\pm$    50&  60$\pm$  91&   3& 1.205e+12$\pm$  93\% \\
  1787.9& -1515.6&  0.31265&    103$\pm$    37&  18$\pm$  27&   5& 1.161e+12$\pm$  76\% \\
   177.5&  2693.2&  0.31704&    357$\pm$   394& 853$\pm$1720&   4& 6.038e+13$\pm$ 152\% \\
   333.2&  -417.7&  0.31759&    272$\pm$   318& 173$\pm$ 260&   3& 8.104e+12$\pm$ 123\% \\
  -384.7& -1033.5&  0.33391&    131$\pm$   117& 141$\pm$ 445&   3& 6.826e+12$\pm$ 219\% \\
  2210.9& -1328.6&  0.35762&    302$\pm$   230& 705$\pm$1242&   4& 5.002e+13$\pm$ 133\% \\
   649.6&  -489.7&  0.35986&    358$\pm$   428&1182$\pm$3100&   3& 4.096e+13$\pm$ 162\% \\
   466.6&  -278.1&  0.37338&    174$\pm$   192& 269$\pm$ 394&   3& 1.064e+13$\pm$ 124\% \\
   750.2&  -472.6&  0.39222&    134$\pm$   107&  49$\pm$ 182&   3& 2.890e+12$\pm$ 213\% \\
  2195.1& -1708.2&  0.39364&    214$\pm$   142& 207$\pm$ 379&   4& 1.986e+13$\pm$ 108\% \\
    -3.6&  2305.5&  0.39380&    153$\pm$   181&  41$\pm$  99&   3& 5.103e+12$\pm$ 140\% \\
   634.7&   241.2&  0.42607&    129$\pm$   104&   7$\pm$  14&   4& 1.492e+12$\pm$ 151\% \\
   517.2&  -474.9&  0.44011&    620$\pm$   314& 915$\pm$ 762&   4& 1.662e+14$\pm$  60\% \\
   827.2& -1016.7&  0.44041&    208$\pm$   212& 134$\pm$ 293&   3& 2.160e+13$\pm$ 166\% \\
   555.6&   305.8&  0.46596&    129$\pm$   109&  25$\pm$  56&   3& 3.308e+12$\pm$ 167\% \\
\hline
\end{tabular}

\newpage

\newcounter{figi}
\newcommand{\nfig}{\addtocounter{figi}{1}\thefigi}

\figcaption[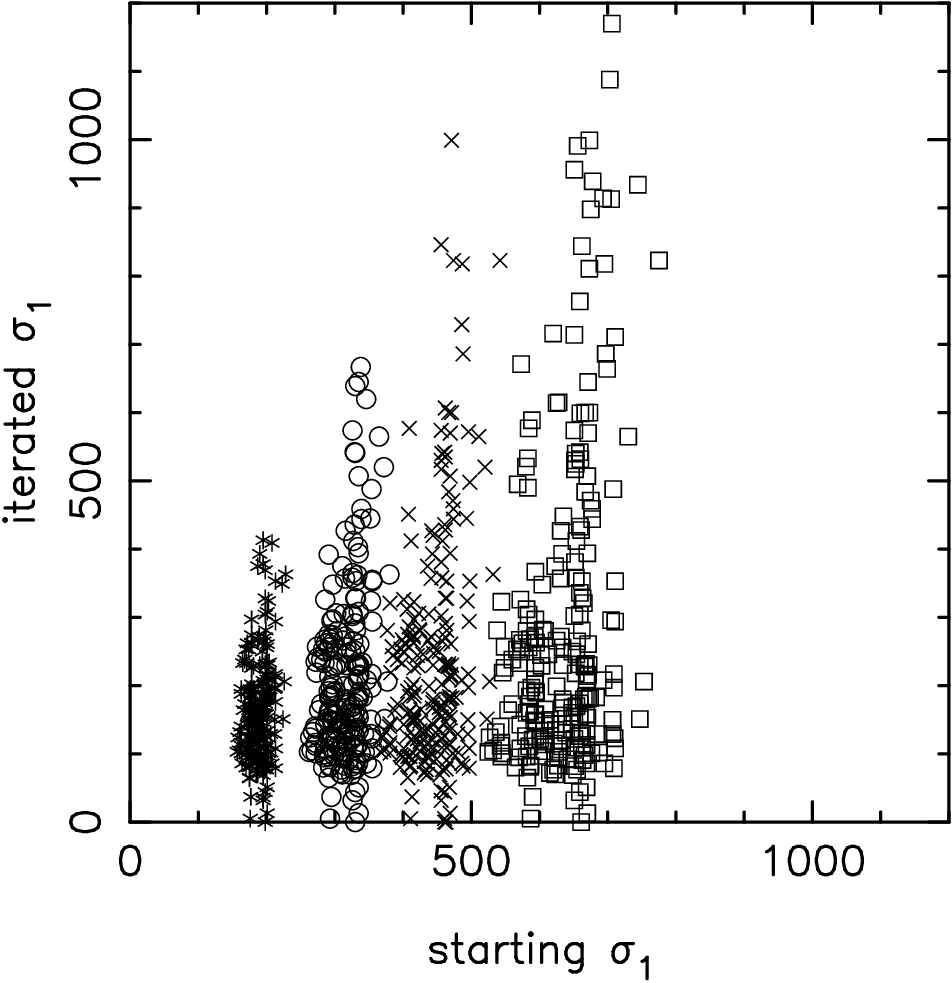]{The line of sight velocity dispersion as
estimated on the first pass (abscissa) of the group refinement and on
the fourth pass (ordinate) for sets of group finding parameters. The
$\rpmax$ is fixed at $0.25\hmpc$. The asterisks, circles, x's and
squares are for $\rzmax$ of 3, 5, 7 and 10\hmpc, respectively.
\label{fig:sig12}}

\figcaption[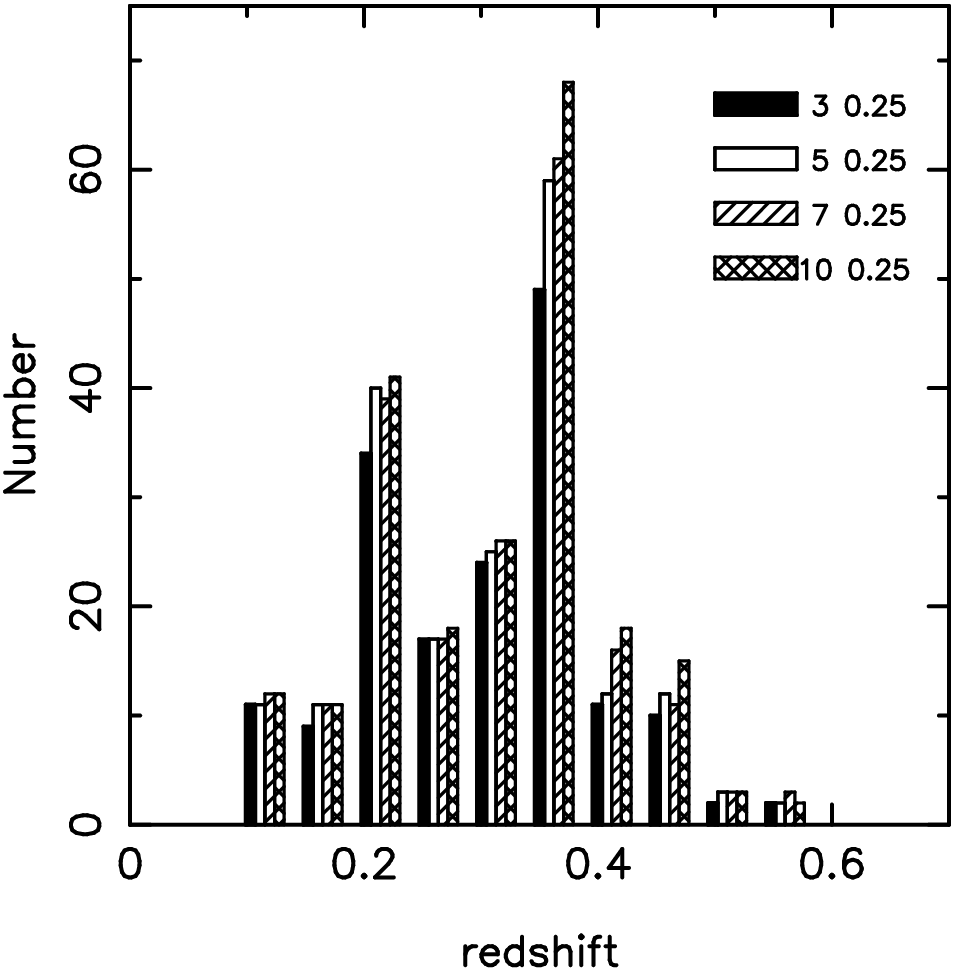]{
The redshift distribution of groups found with $r_p^{max}$ fixed at
$0.25\hmpc$ and $r_z^{max}$ (indicated on the figure in the upper
right) varying over the same range as Figure~\ref{fig:sig12} .
\label{fig:nz}}

\figcaption[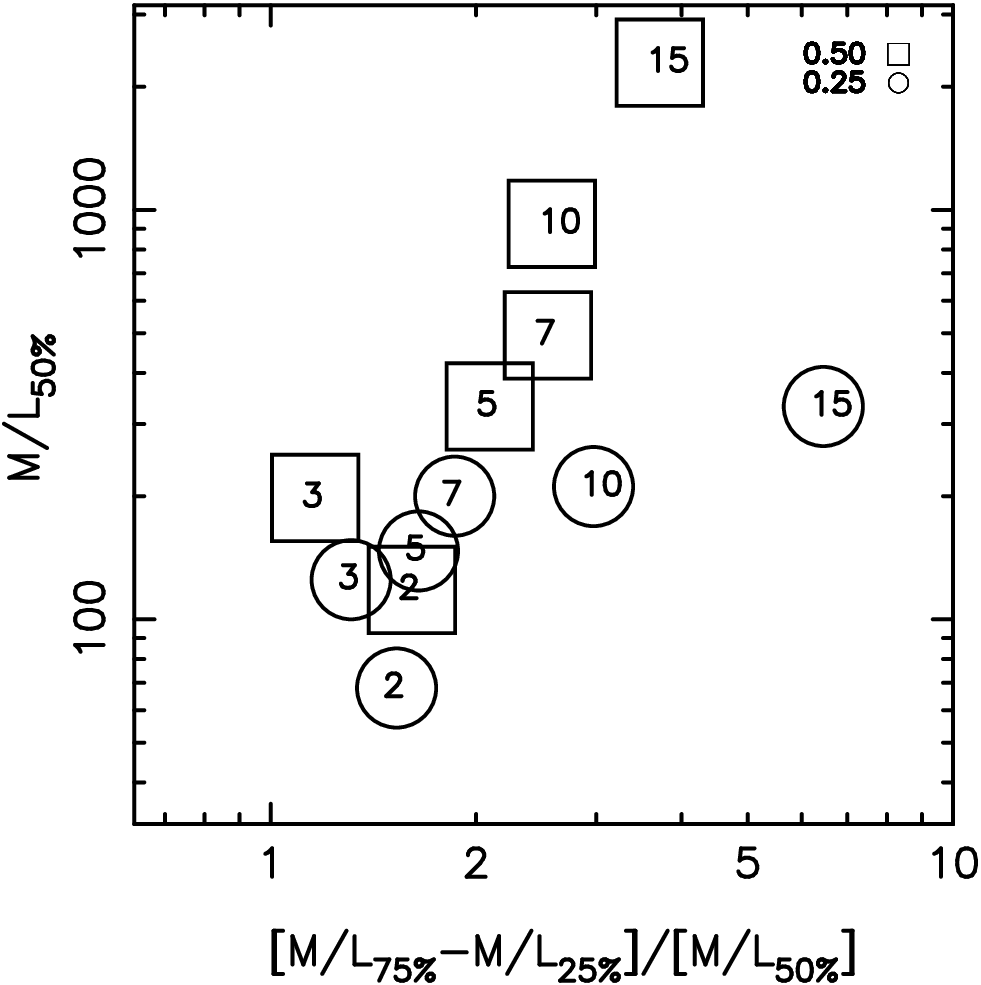]{
The median $M/L$ value of the groups versus the factional spread
between the first and third quartile $M/L$ values.  Squares indicate
$\rpmax=0.5\hmpc$ groups and circles indicate $\rpmax=0.25\hmpc$
groups. The numbers in the symbols give the values of $\rzmax$.
\label{fig:mls}}

\figcaption[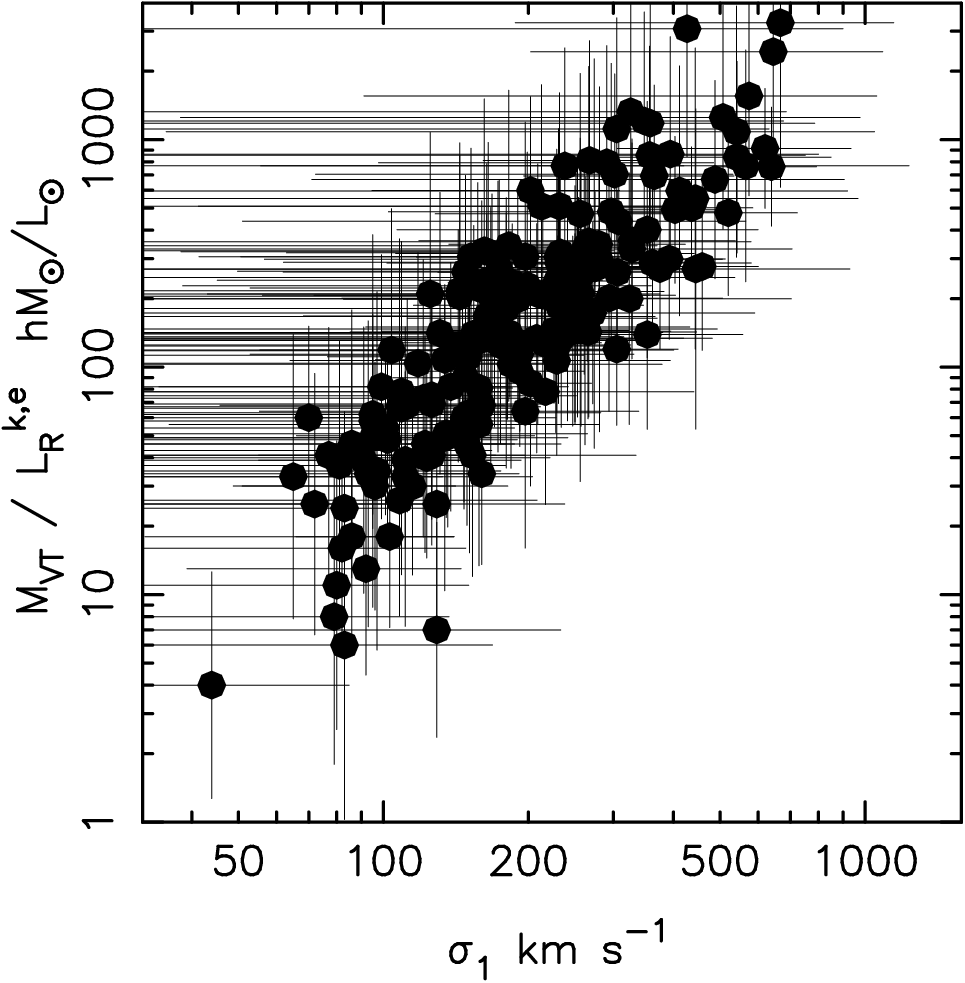]{ 
The mass-to-light ratios of the groups as a function of their velocity
dispersions for $\rpmax=0.25\hmpc$ and $\rzmax=5\hmpc$. The $1\sigma$
error flags plotted are Jackknife technique estimates. The correlation
is the result of the $\sigma_1^2$ dependence of $M_{VT}$. The inset
restricts the sample to groups with six or more members.
\label{fig:ml_sig}}

\figcaption[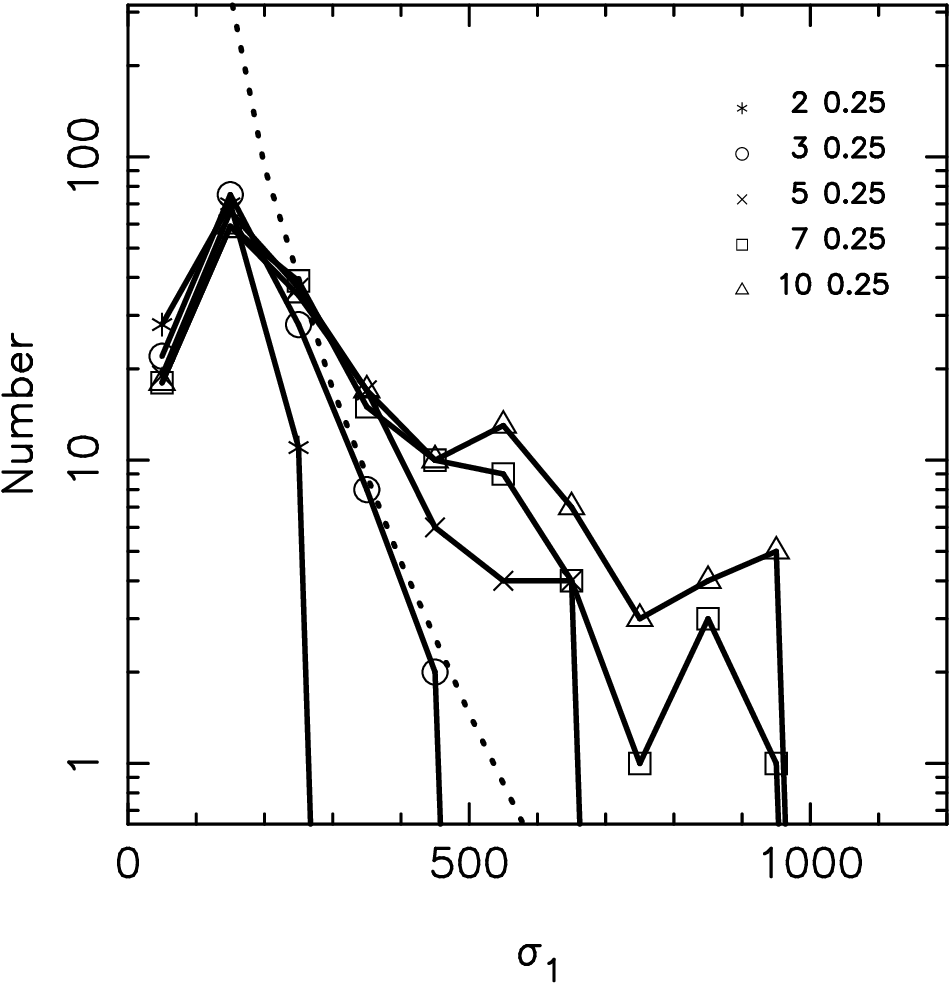]{The distribution of measured line-of-sight
velocity dispersions in 100\kms\ bins, plotted at the bin centers.
The parameters for the group selection are indicated by the symbols in
the diagram.  The dotted line shows the Press-Schechter prediction for
the distribution using the cluster normalization.
\label{fig:lnsig}}

\figcaption[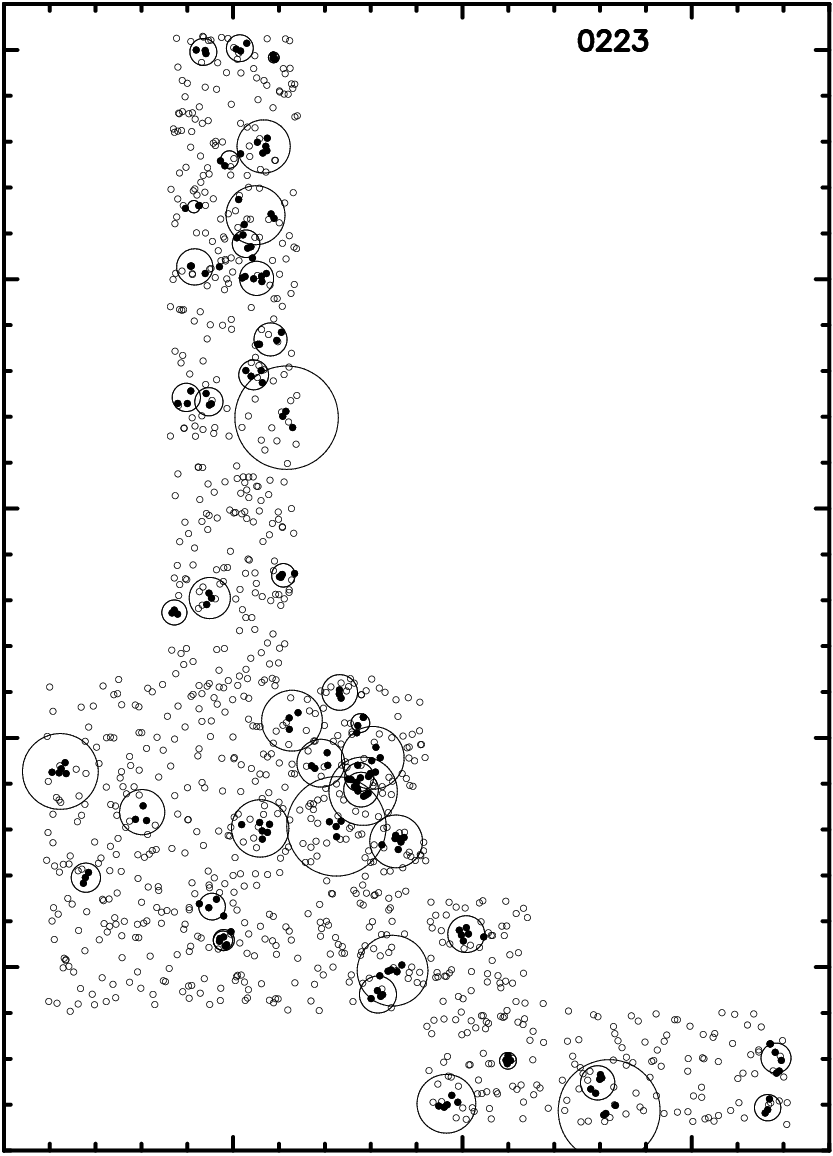]{
The $xy$ locations of field galaxies (open circles) and group
galaxies (filled), along with $r_{200}$ radii of the groups found between
redshifts 0.1 and 0.45 in the 0223+00 field. The starting parameters
are $r_p^{max}=0.25\hmpc$ and $r_z^{max}=5\hmpc$ (co-moving). The
group members are marked with filled symbols. Note that the group
members fill their $r_{200}$ to varying degrees. None of the members
are outside the estimated region of virialization estimated for a low
density universe, $1.5r_{200}$. The major ticks are at intervals
of 1000 arcsec, with North up and East to the left.
\label{fig:xy}}

\figcaption[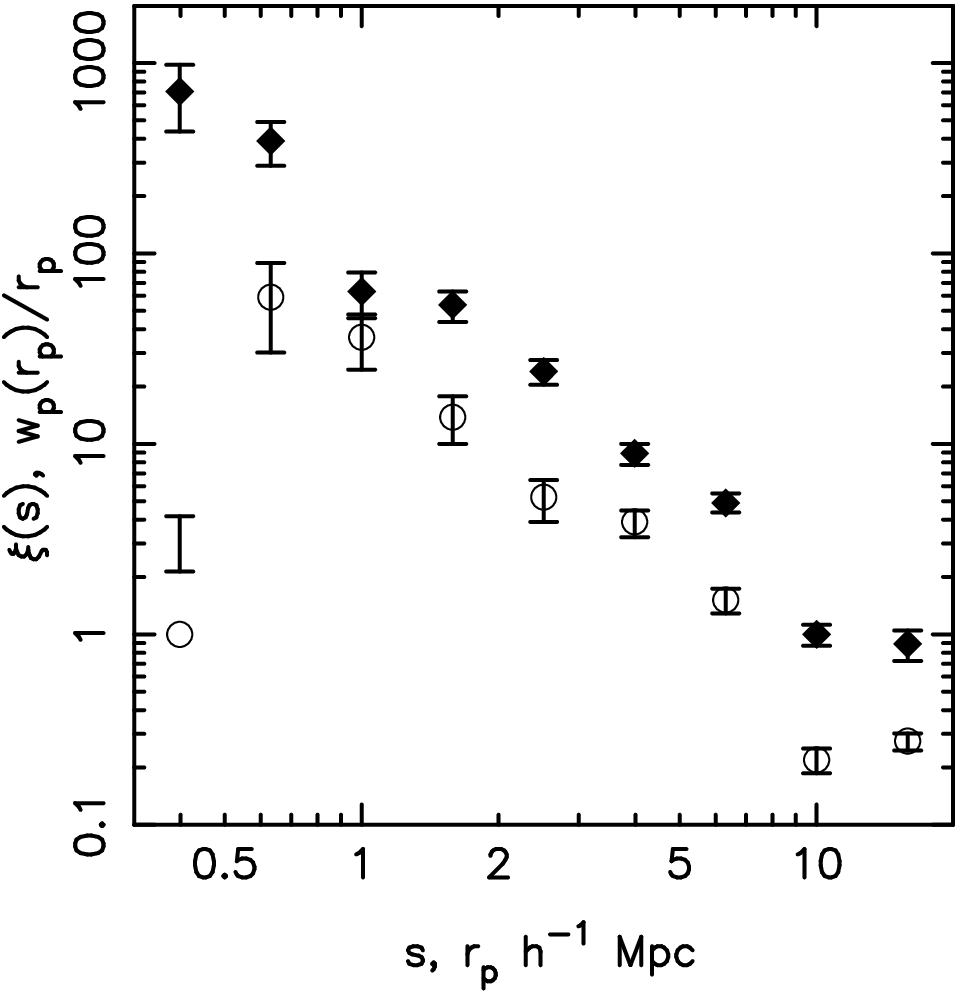]{The redshift space (open circles) and projected real space
(filled diamonds) correlations for the $r_p^{max}=0.25\hmpc$ and
$r_z^{max}=5\hmpc$ groups. The errors here are simply the square
root of the pair count in the bin. The fitted correlation lengths are
$6.8\pm 0.3\hmpc$ and $6.5\pm0.3\hmpc$, respectively.
\label{fig:gpxi}}

\figcaption[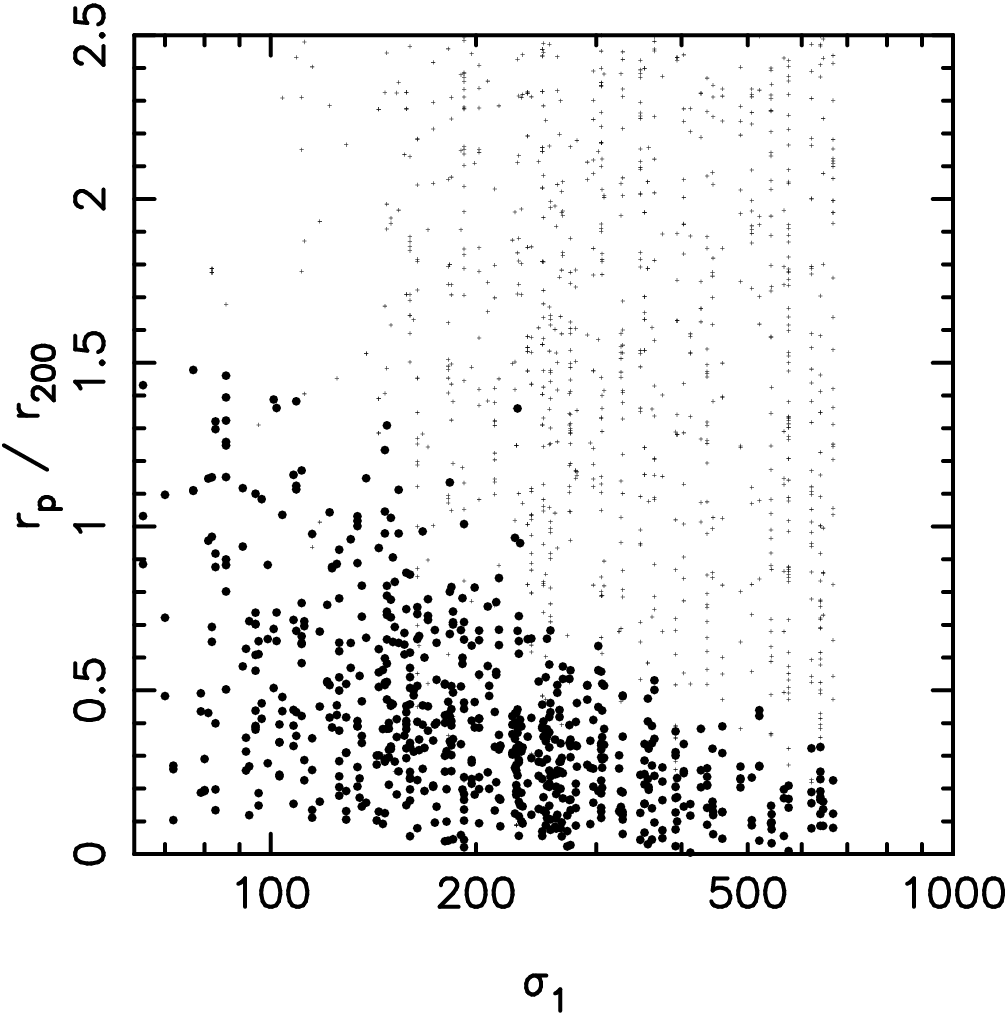]{
The scaled radius versus the source group's velocity dispersion for
all galaxies in the mean group for $\dvmax=3\sigma_1$ prior to
statistical background subtraction. The small plus signs are for all
galaxies contributing to the mean group whereas the filled octagons
mark the galaxies identified as group members using our algorithm.
\label{fig:rr200}}

\figcaption[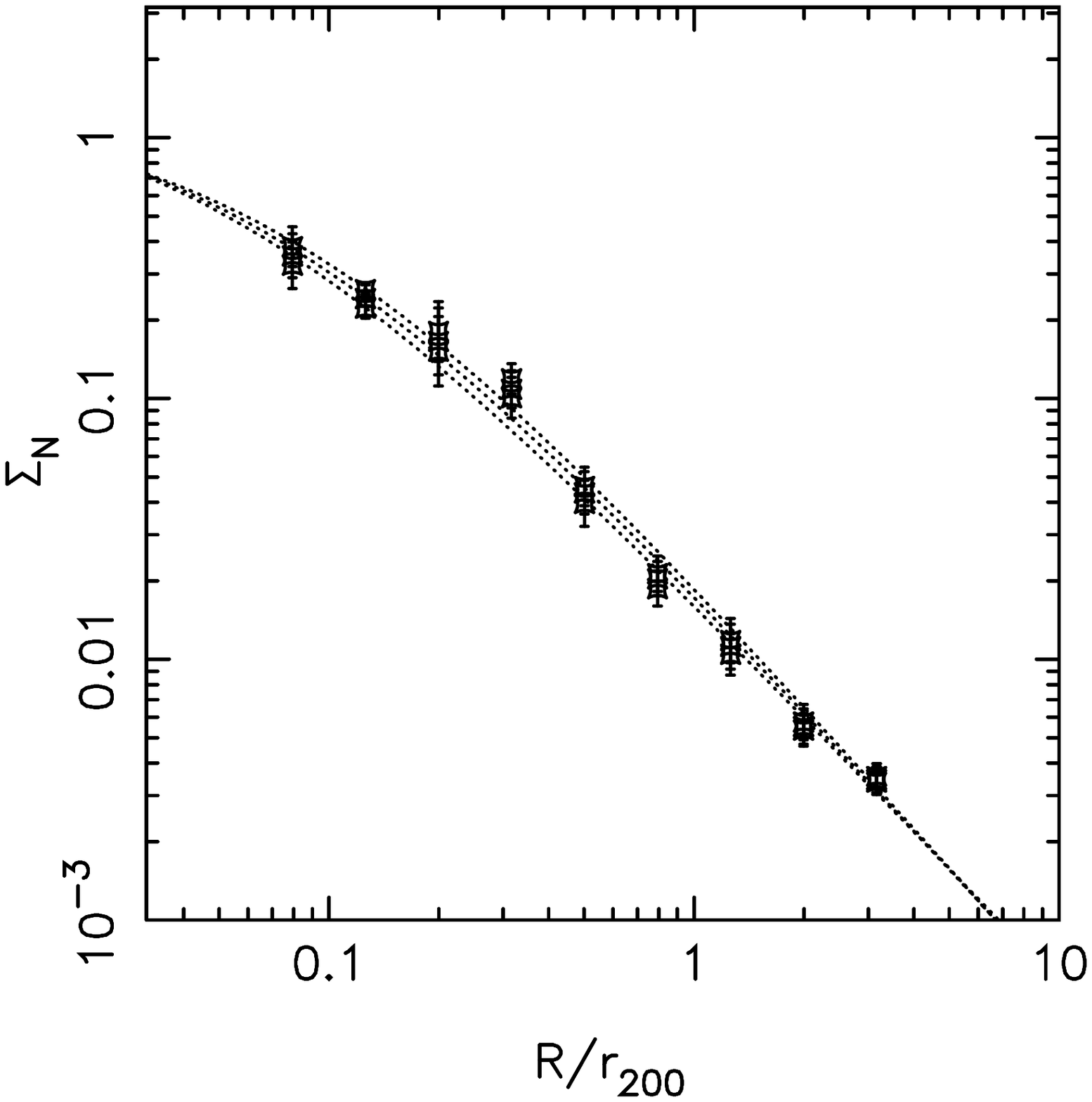]{The number weighted surface density of
galaxies in the mean group. The points and three fitted lines show the
slow {\it decrease} of the normalized surface density at intermediate
radii with clipping at 3, 4 and 5 velocity units.  The inset shows
$\Sigma_N(R)$ using only the $\sigma_1\le 200\kms$ groups with (solid
line) and, for reference, all velocity dispersion groups cutoff again
(dotted line), both with 3 velocity unit clipping. 
\label{fig:osurf}}

\figcaption[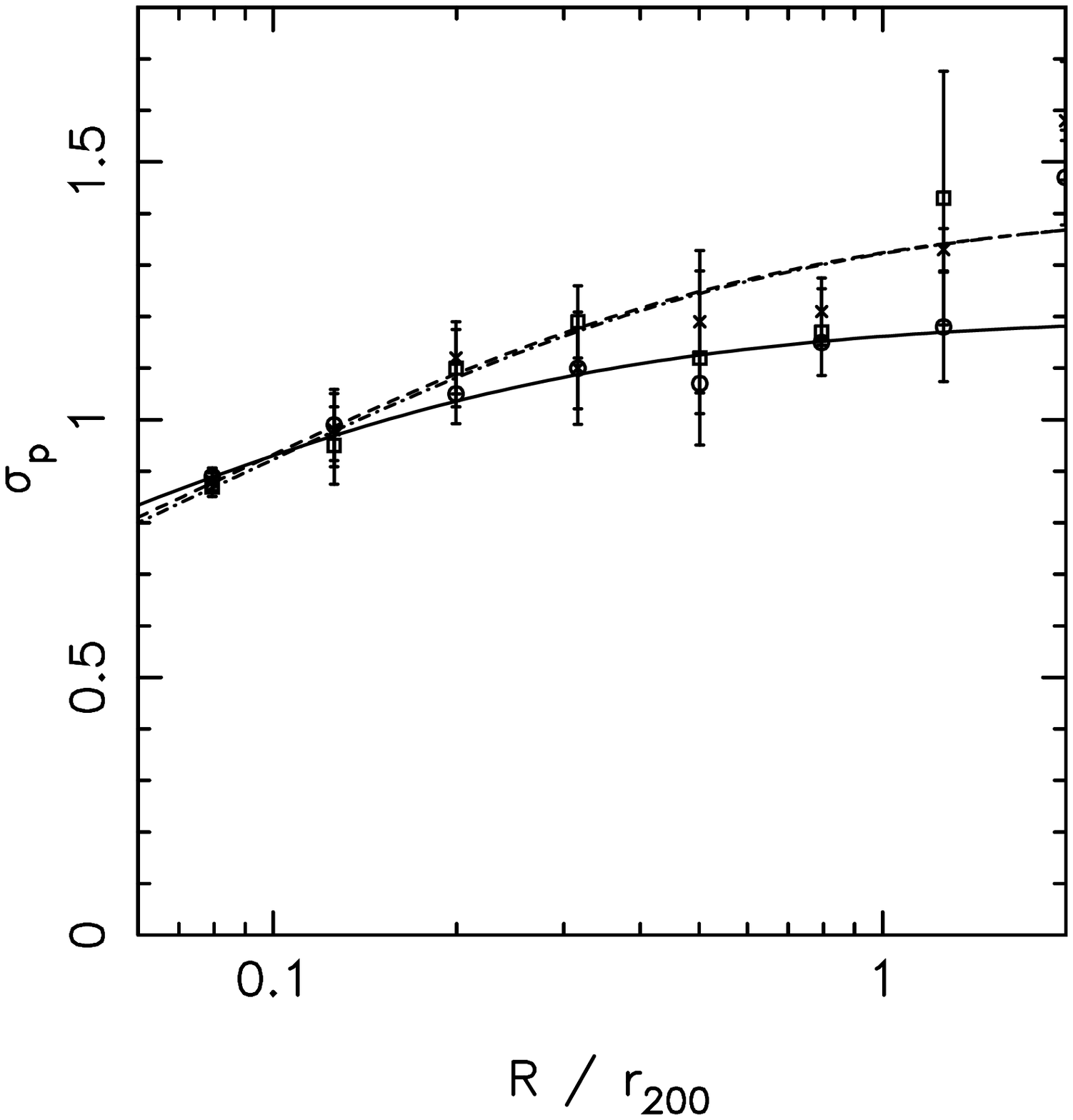]{The number-weighted projected velocity dispersion
profile. The circle, cross and square are for velocity cuts at 3, 4
and 5 velocity unit cutoff, which are fit by the solid line and the
dashed and dotted lines (which are hard to distinguish because they
are overlaid).
\label{fig:osig}}

\figcaption[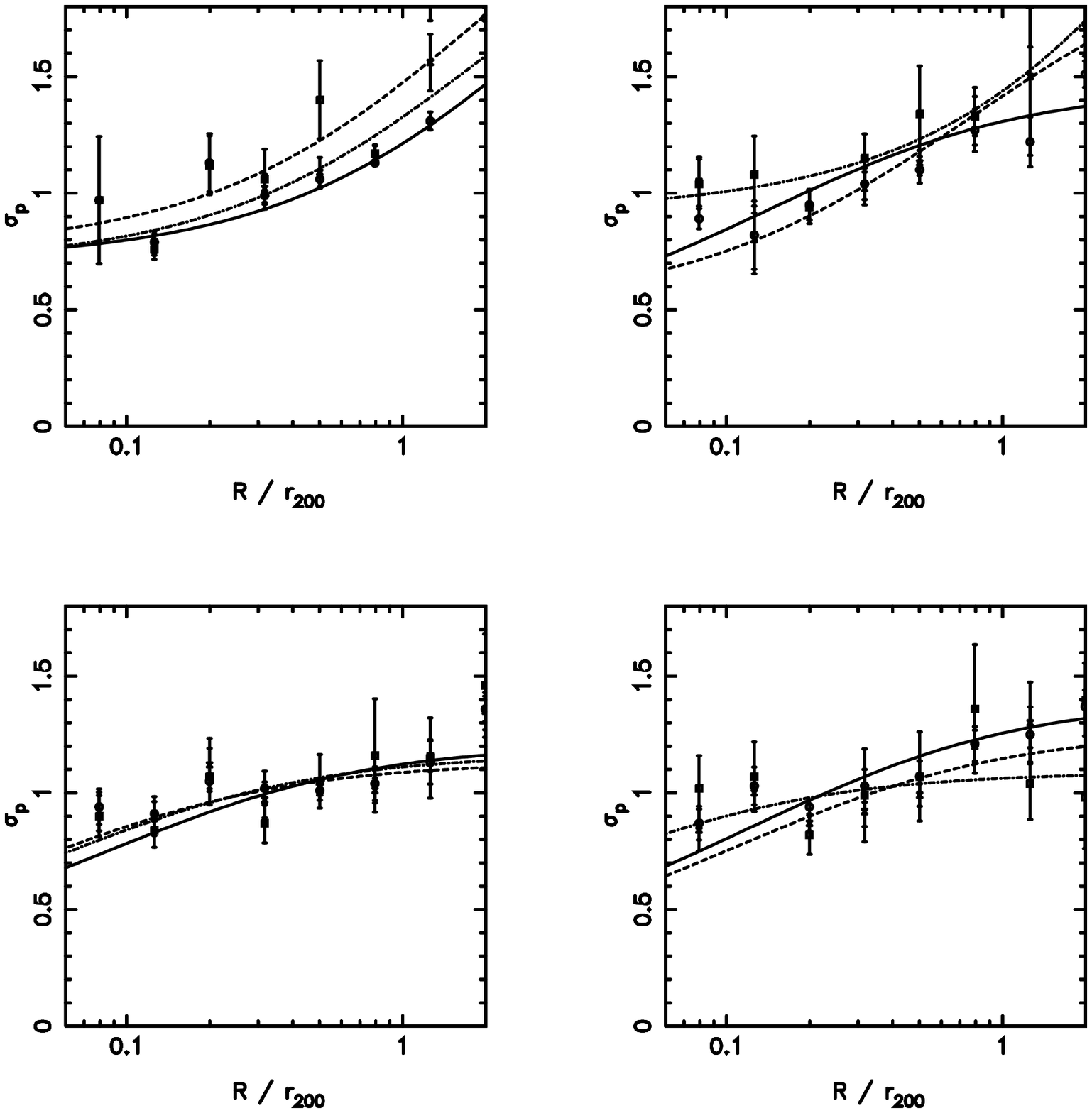]{The number-weighted projected velocity dispersion
profile for alternate values of $\rzmax$ of 2, 3, 7 and 10 \hmpc. The
circle, cross and square are for velocity cuts at 3, 4 and 5 velocity
units, and fitted by solid, dashed and dotted lines, respectively.
\label{fig:osig_rz}}

\figcaption[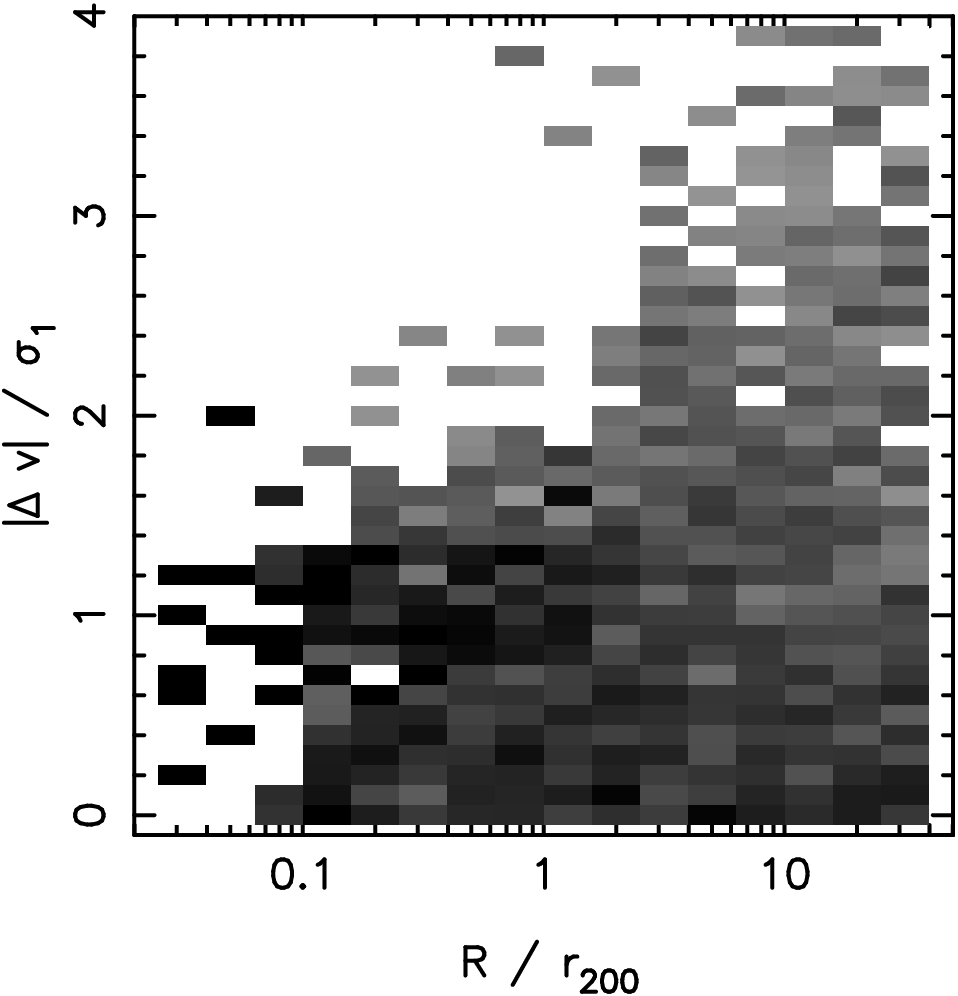]{The mean radially normalized
group in position and velocity, $\xi(R,\Delta v)/\Sigma(r_p)$,
selected using $\rpmax=0.25\hmpc$ and $\rzmax=5\hmpc$.  Each column of
fixed $R$ is normalized to a total of unity. The gray scale is
proportional to the logarithm of the local over-density in the
normalized velocity and radius.  The second moment of this
distribution in the vertical direction is $\sigma_p(R)$ of
Figure~\ref{fig:osig}.  The lowest over-density value (white) is 0.2\%
of the total, and the highest (black) is 10\% of the total in the
entire $R$ bin. 
\label{fig:gray}}

\figcaption[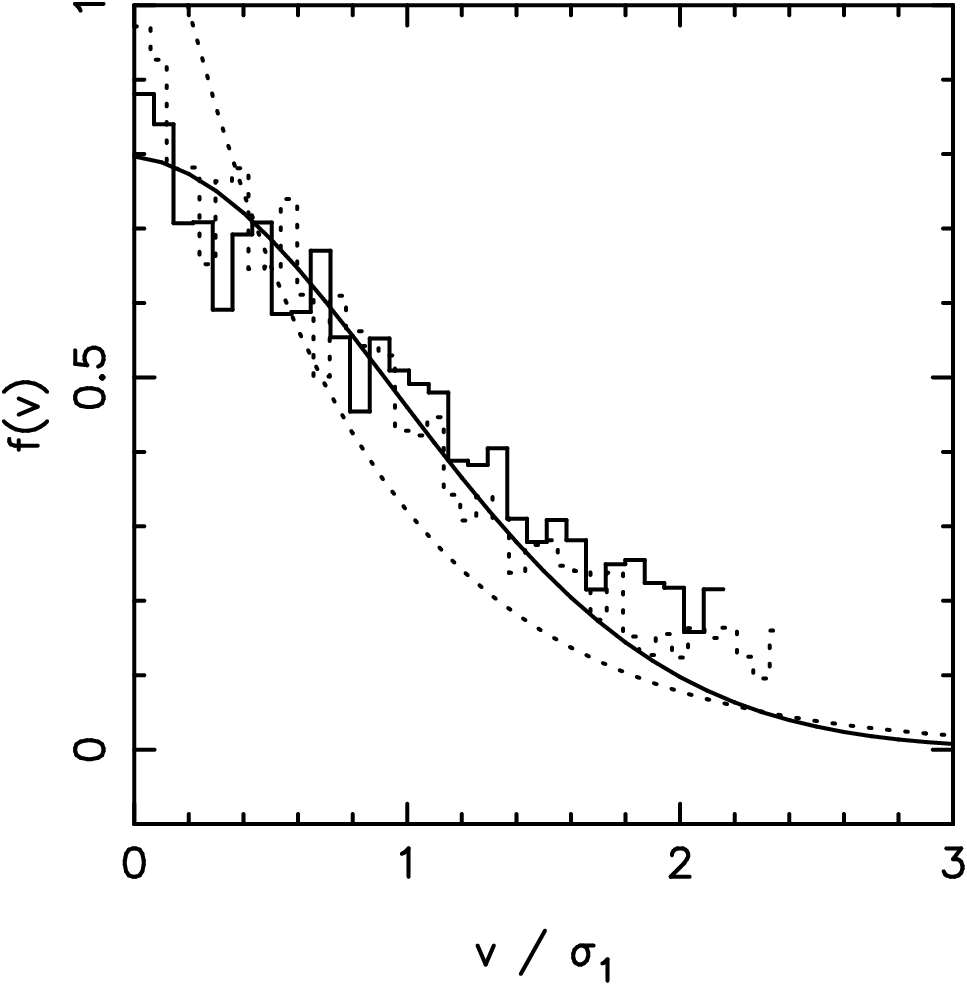]{The projected velocity distribution function (pvdf), 
normalized to have unit velocity dispersion and unit area under the
curve.  The pvdf for cutoffs of 3 and 4 velocity units are shown as
solid and short dashed stepped lines, respectively.  The reference
lines are a Gaussian and an exponential distribution, solid and
short-dashed lines, respectively.
\label{fig:vdf}}

\figcaption[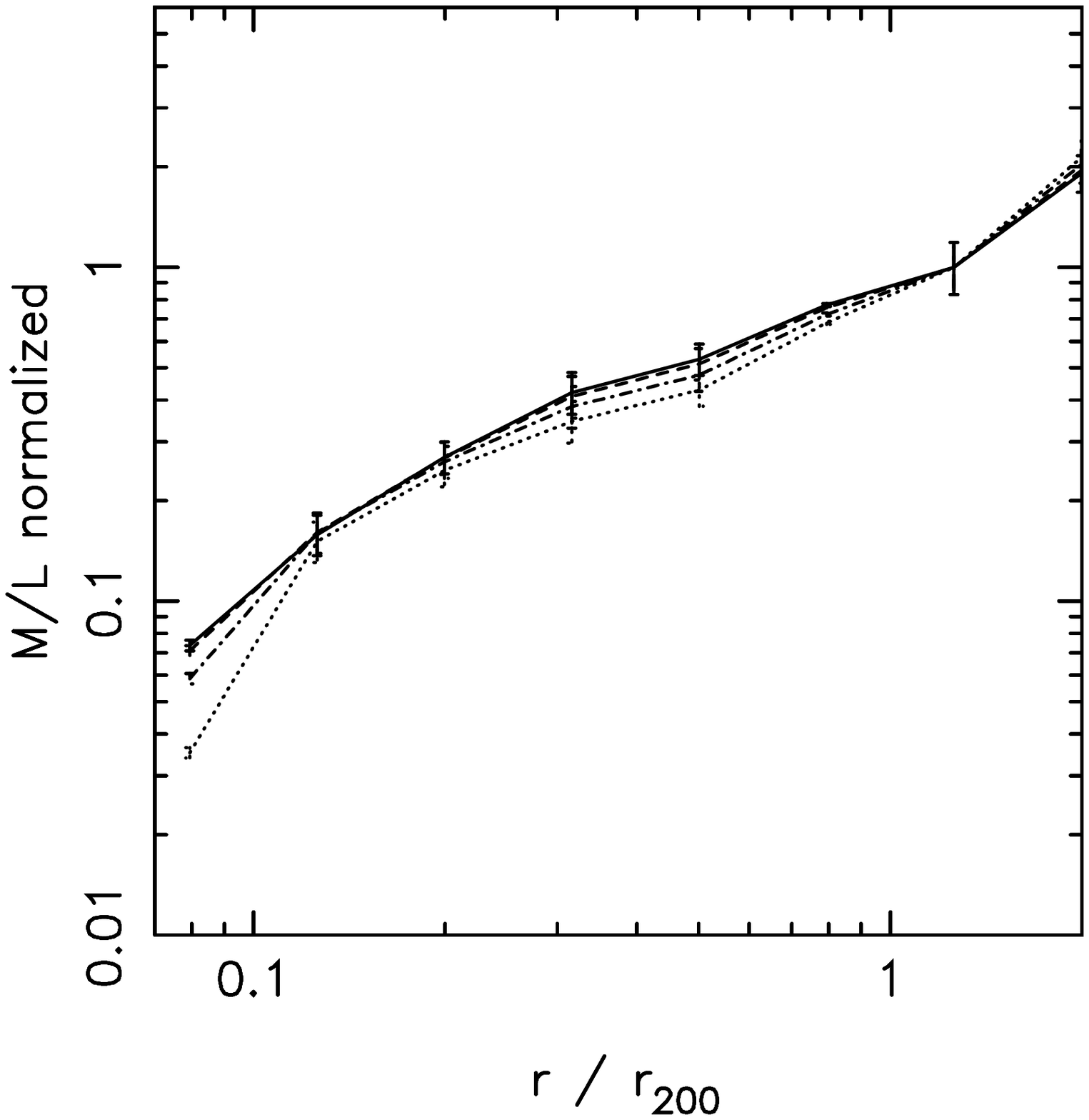]{The rising mass-to-light ratio from
the 3 velocity unit cutoff velocity dispersion profile for group
centers identified with $r_p^{max}=0.25\hmpc $ and
$r_z^{max}=5\hmpc$. The galaxy co-ordinates are scaled to their
groups $\sigma_1$ and $r_{200}$. The errors are computed from the
velocity dispersion errors. The upper inset shows the results obtained
by overlaying the galaxies in physical co-ordinates assuming
$\sigma_1=200\kms$ and the lower shows the results after restricting
the analysis to groups with velocity dispersion less than 200
\kms. The solid, dashed, dot-dashed and dotted lines are for
$\beta_\infty=-1,-\onehalf ,0$, and $\onequarter$, respectively
\label{fig:mol}}

\figcaption[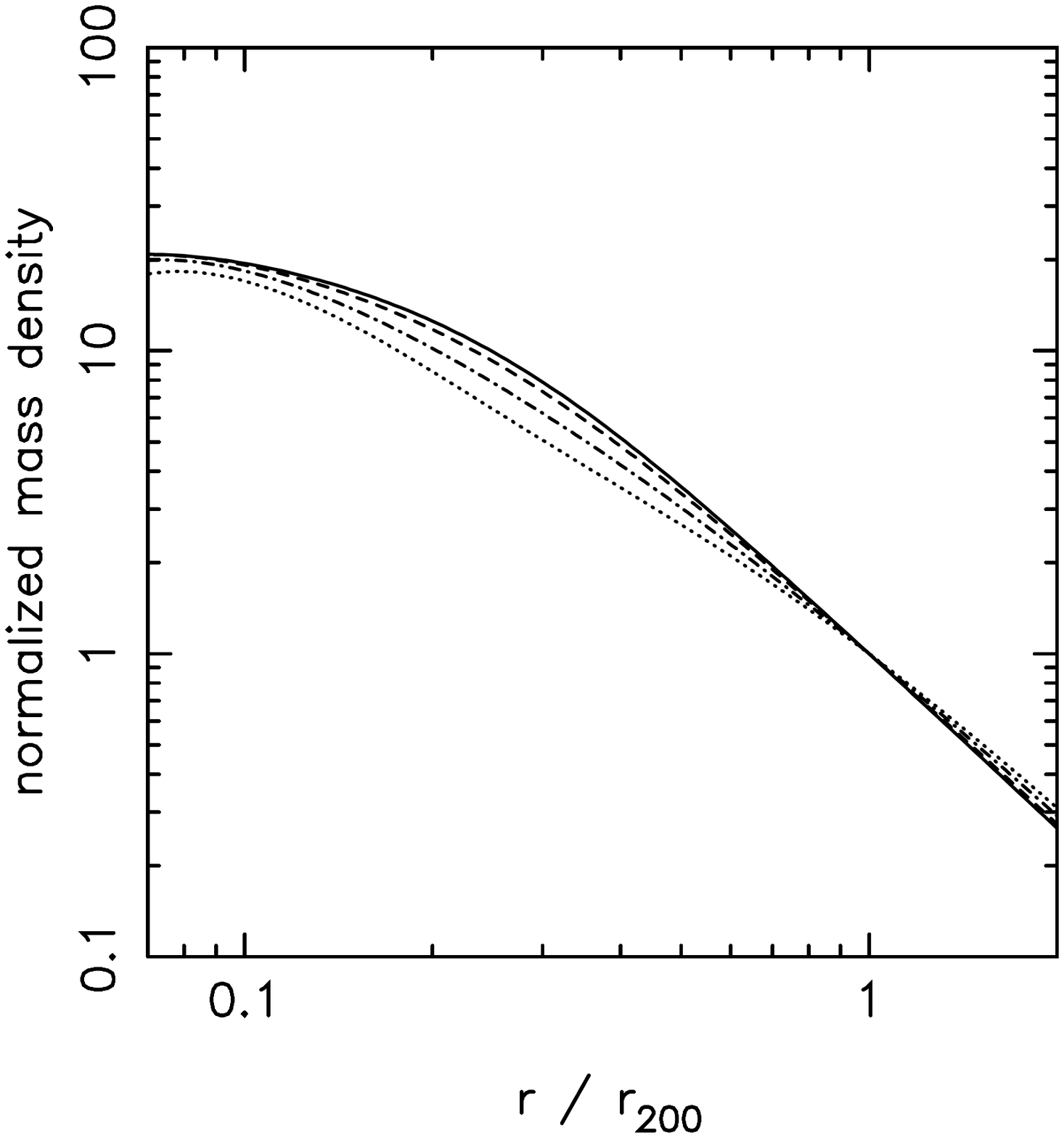]{The inferred dark matter density profile for
the variable velocity anisotropy model, with $\beta_0=0$,
$r_\beta=0.3r_{200}$ and $\beta_\infty=-1,-\onehalf,0$ and
$\onequarter$ for solid, dashed, dot-dashed and dotted lines,
respectively, as in Figure~\ref{fig:mol}.  The lower inset
shows the results obtained by overlaying the galaxies in physical
co-ordinates and the upper shows the results after restricting the
analysis to groups with velocity dispersion less than 200 \kms.
\label{fig:den}}

\figcaption[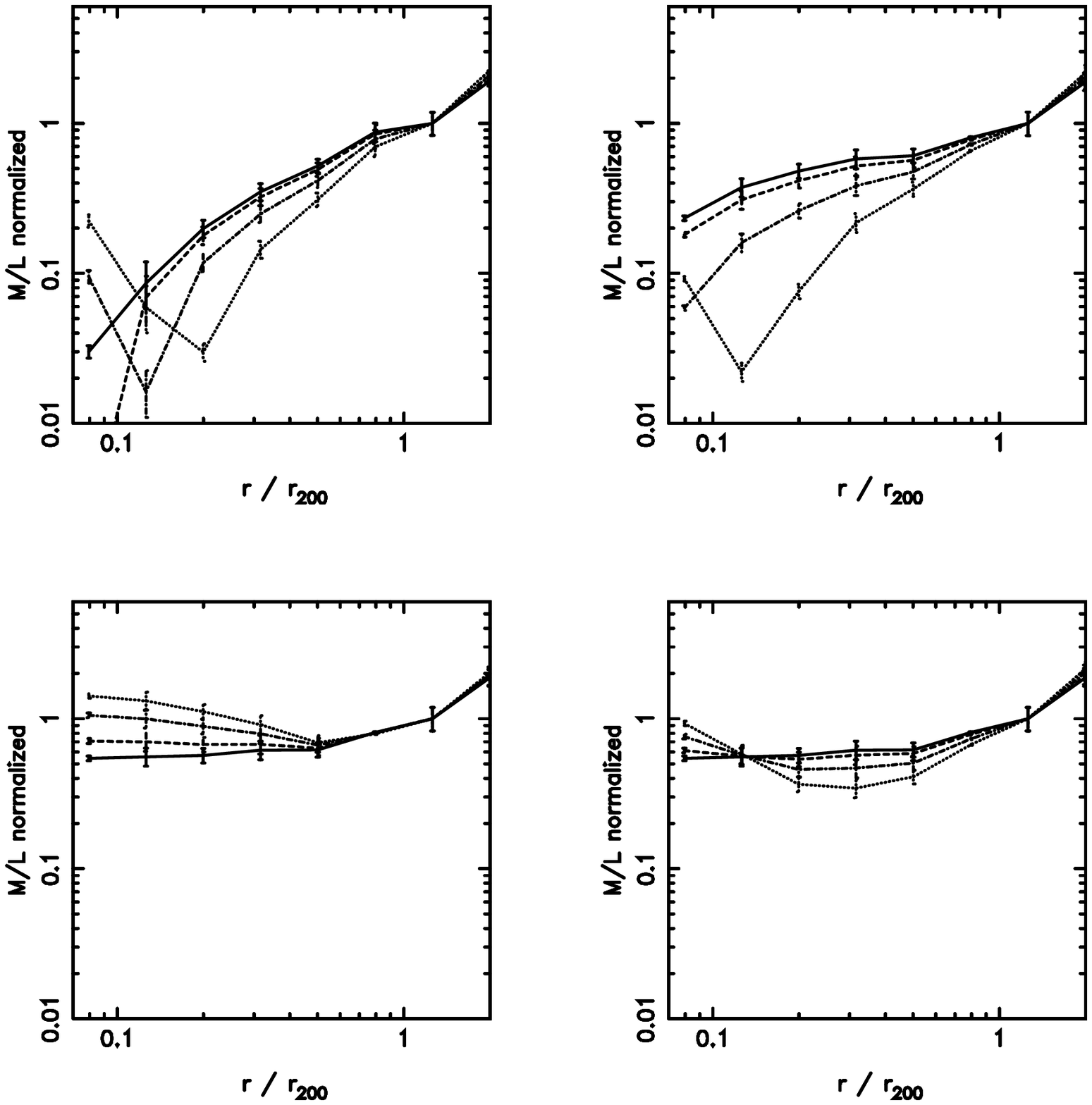]{Profiles of $M/L$ for an alternate
data set, $\rzmax=3\hmpc$ and $\rpmax=0.25\hmpc$ (upper left) and the
standard groups sample analyzed with a small $r_\beta=0.1r_{200}$
(upper right). In both lower panels we use an extremely tangential
velocity ellipsoid, $\beta_0=-1$ with the standard
$r_\beta=0.3r_{200}$ (lower left), and a small $r_\beta=0.1r_{200}$,
(lower right). The lines have the same meaning as in
Figure~\ref{fig:den}. When the $M/L$ has a zero-crossing dip the the
implied values of the mass are negative at small radii.  These show
that a nearly tangential velocity ellipsoid can eliminate the inference
of a rising mass-to-light ratio. 
\label{fig:omol}}

\figcaption[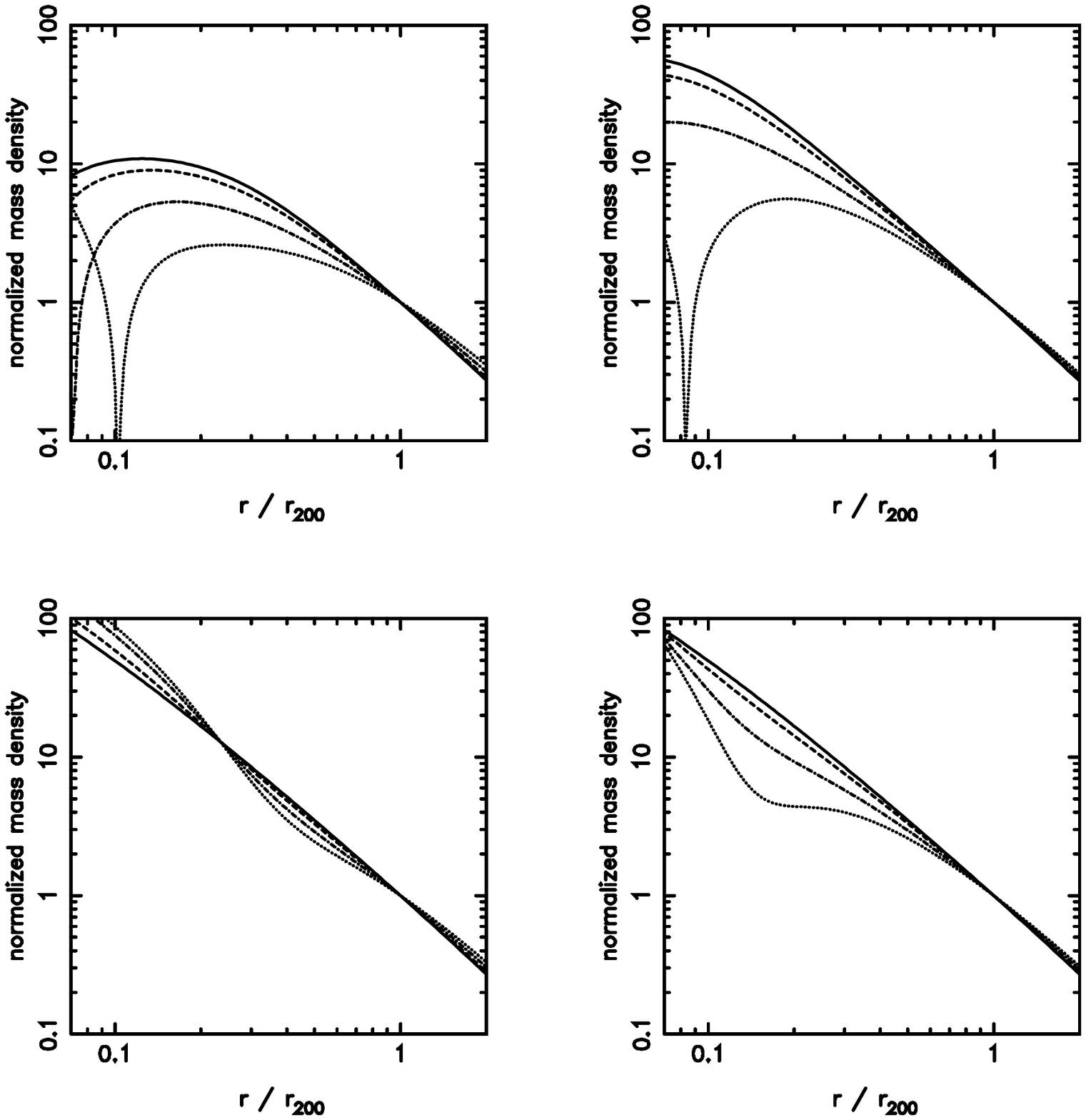]{The density profiles for
the data and models of Figure~\ref{fig:omol}. Again, masses are
negative at small radii when a zero-crossing dip is present.
\label{fig:oden}}

\figcaption[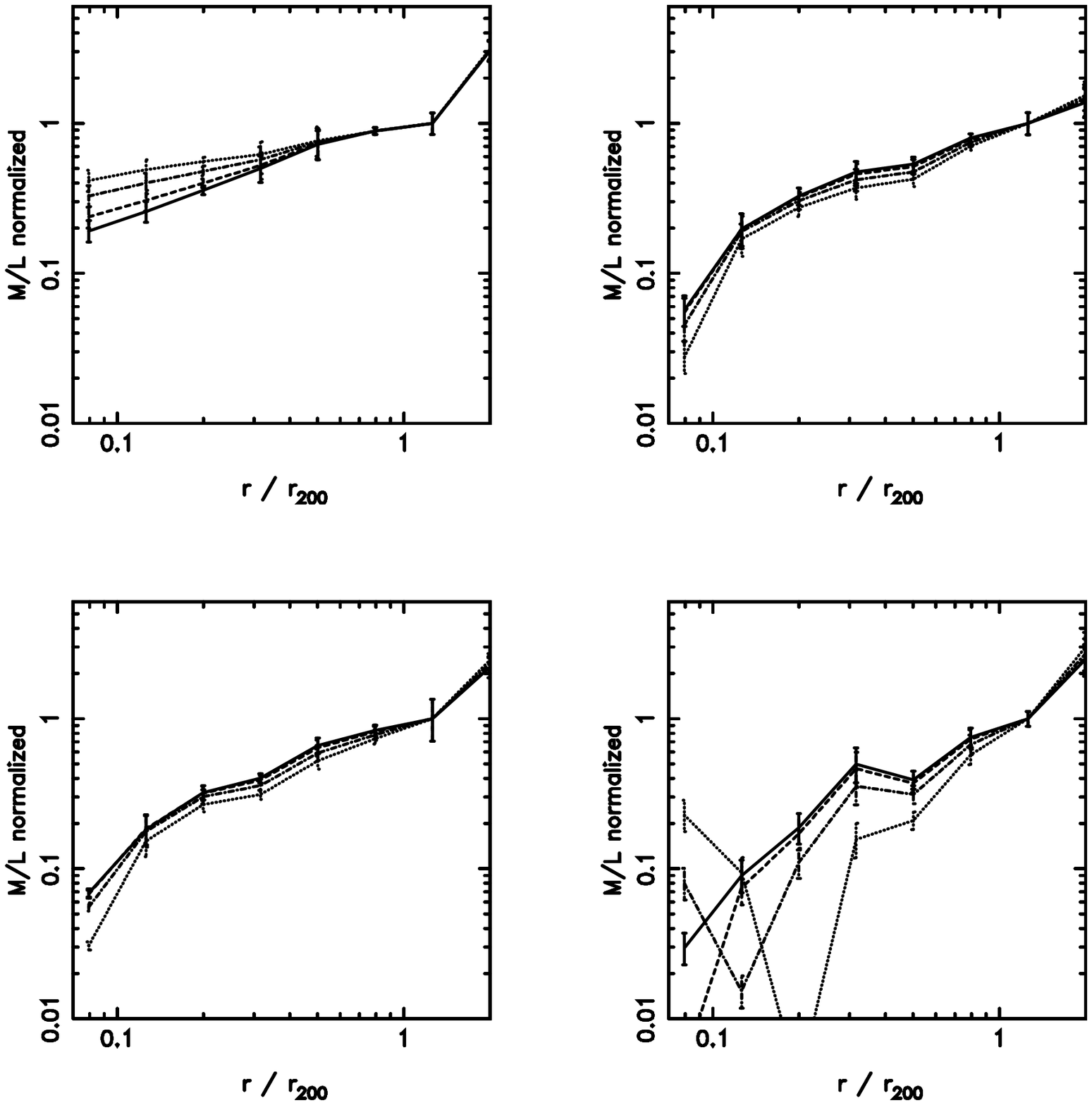]{The inferred mass-to-light ratios (normalized
to unity at $r_{200}$) for galaxy subsamples within the groups. The
top row has blue (left) and red (right) selected galaxies. The bottom
row has high (left) and low (right) luminosity galaxies. The smooth
lines show the modeled results. The main result is that blue galaxies
trace the mass distribution much more closely than the red galaxies.
\label{fig:omol_sub}}

\begin{figure}\figurenum{\nfig}
	\includegraphics[width=0.9\hsize]{fig1.ps} 
\caption{}\end{figure}  

\begin{figure}\figurenum{\nfig}
	\includegraphics[width=0.9\hsize]{fig2.ps} 
\caption{}
\end{figure}  

\begin{figure}\figurenum{\nfig}
	\includegraphics[width=0.9\hsize]{fig3.ps} 
\caption{}\end{figure}  

\begin{figure}\figurenum{\nfig}
\begin{picture}(0,0)(0,0)
        \put(  0,-550){\includegraphics[width=0.9\hsize]{fig4.ps} }
        \put(260,-470){\includegraphics[width=0.3\hsize]{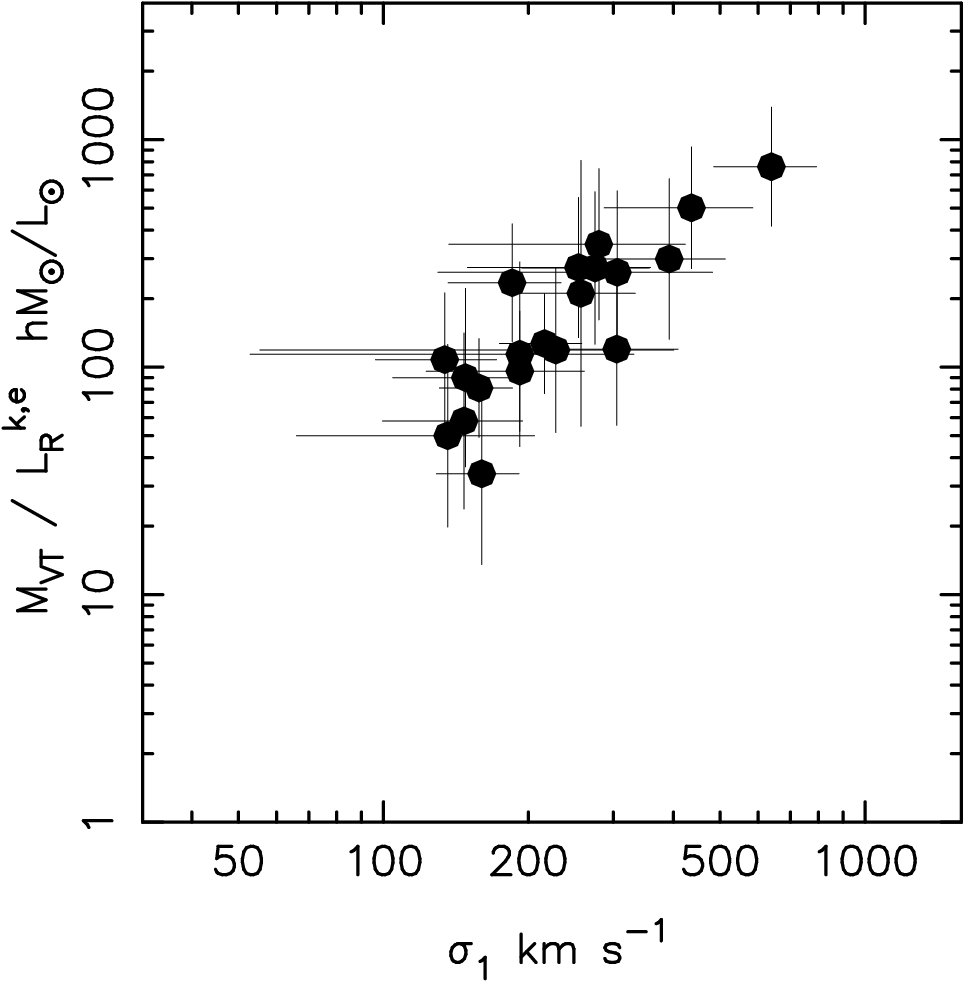}}
\end{picture}
\vspace{20.0cm}
\caption{}\end{figure}  

\begin{figure}\figurenum{\nfig}
	\includegraphics[width=0.9\hsize]{fig5.ps} 
\caption{}\end{figure}  

\begin{figure}\figurenum{\nfig}
	\includegraphics{fig6.ps} 
\caption{}
\end{figure}  

\begin{figure}\figurenum{\nfig}
	\includegraphics[width=0.9\hsize]{fig7.ps} 
\caption{}\end{figure}  

\begin{figure}\figurenum{\nfig}
	\includegraphics[width=0.9\hsize]{fig8.ps}
\caption{}
\end{figure}  

\begin{figure}\figurenum{\nfig}
\begin{picture}(0,0)(0,0)
        \put(  0,-550){\includegraphics[width=0.9\hsize]{fig9.ps} }
        \put(220,-325){\includegraphics[width=0.4\hsize]{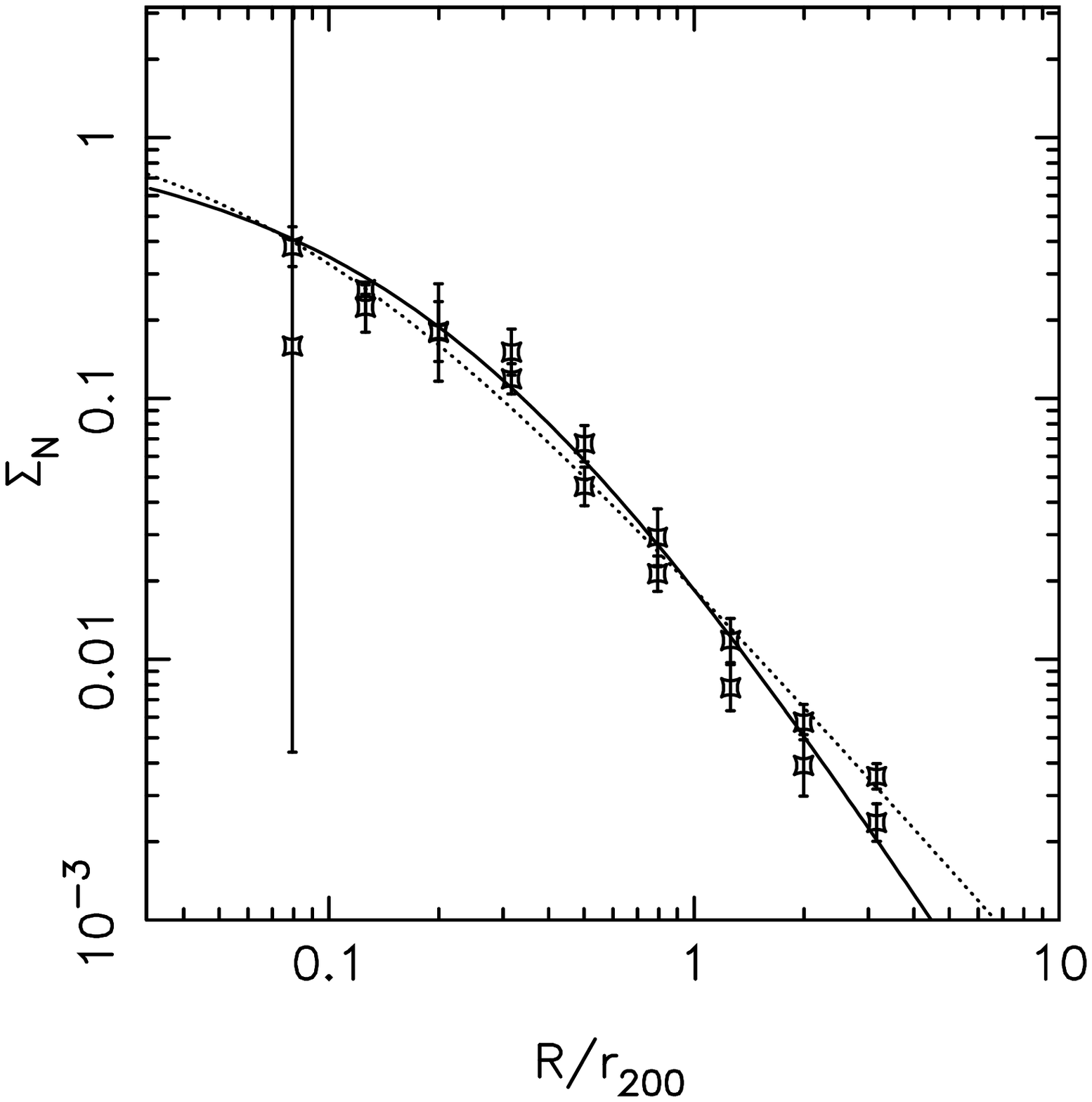}}
\end{picture}
\vspace{20.0cm}
\caption{}\end{figure}  
\clearpage

\begin{figure}\figurenum{\nfig}
	\includegraphics[width=0.9\hsize]{fig10.ps} 
\caption{}\end{figure}  

\begin{figure}\figurenum{\nfig}
	\includegraphics[width=0.9\hsize]{fig11.ps} 
\caption{}\end{figure}  

\begin{figure}\figurenum{\nfig}
        \includegraphics[width=0.9\hsize]{fig12.ps}
\caption{}\end{figure}  

\begin{figure}\figurenum{\nfig}
	\includegraphics[width=0.9\hsize]{fig13.ps} 
\caption{}\end{figure}  

\begin{figure}\figurenum{\nfig}
\begin{picture}(0,0)(0,0)
        \put(  0,-550){\includegraphics[width=0.9\hsize]{fig14.ps} }
        \put( 80,-320){\includegraphics[width=0.3\hsize]{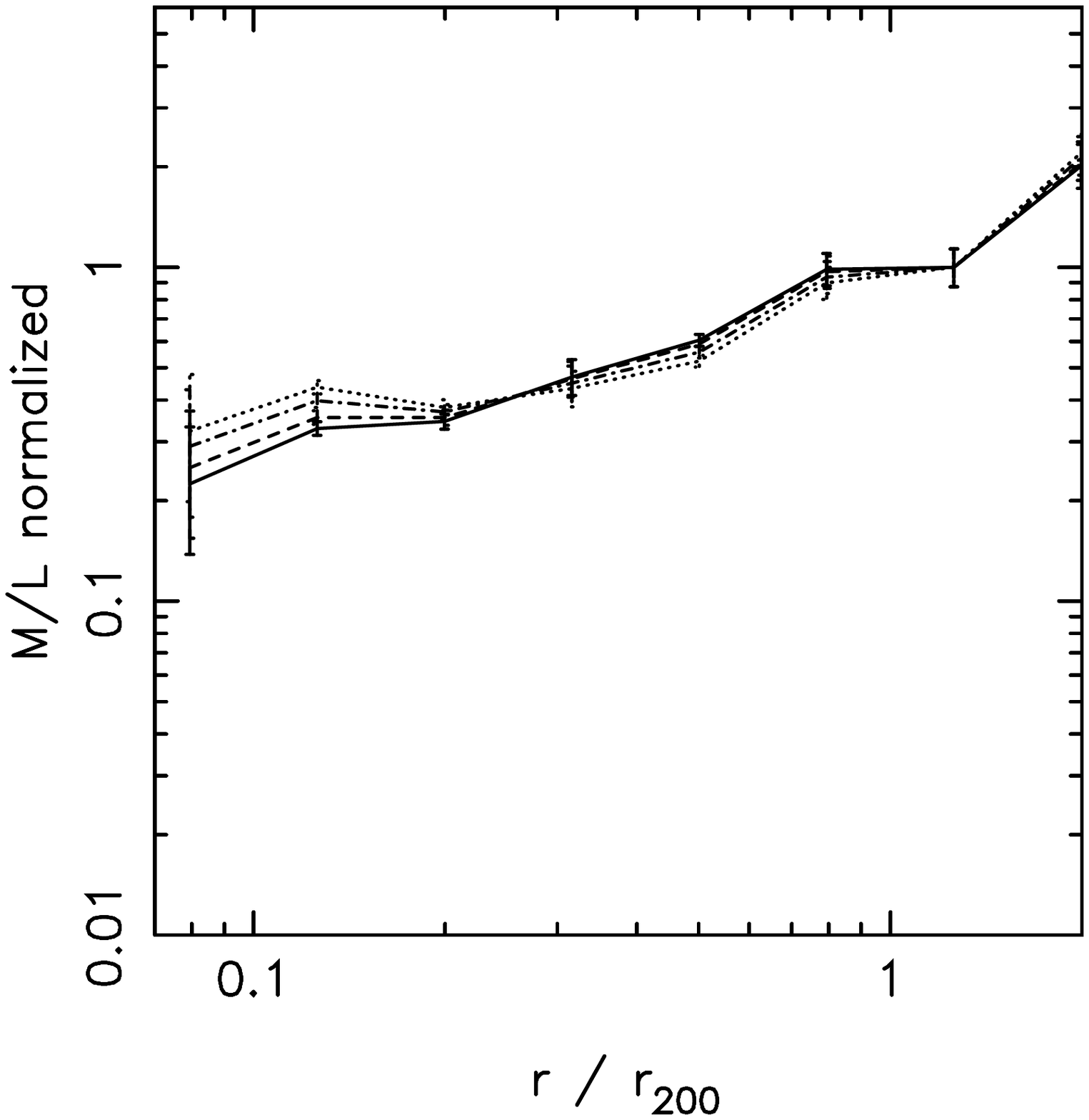}}
        \put(260,-450){\includegraphics[width=0.3\hsize]{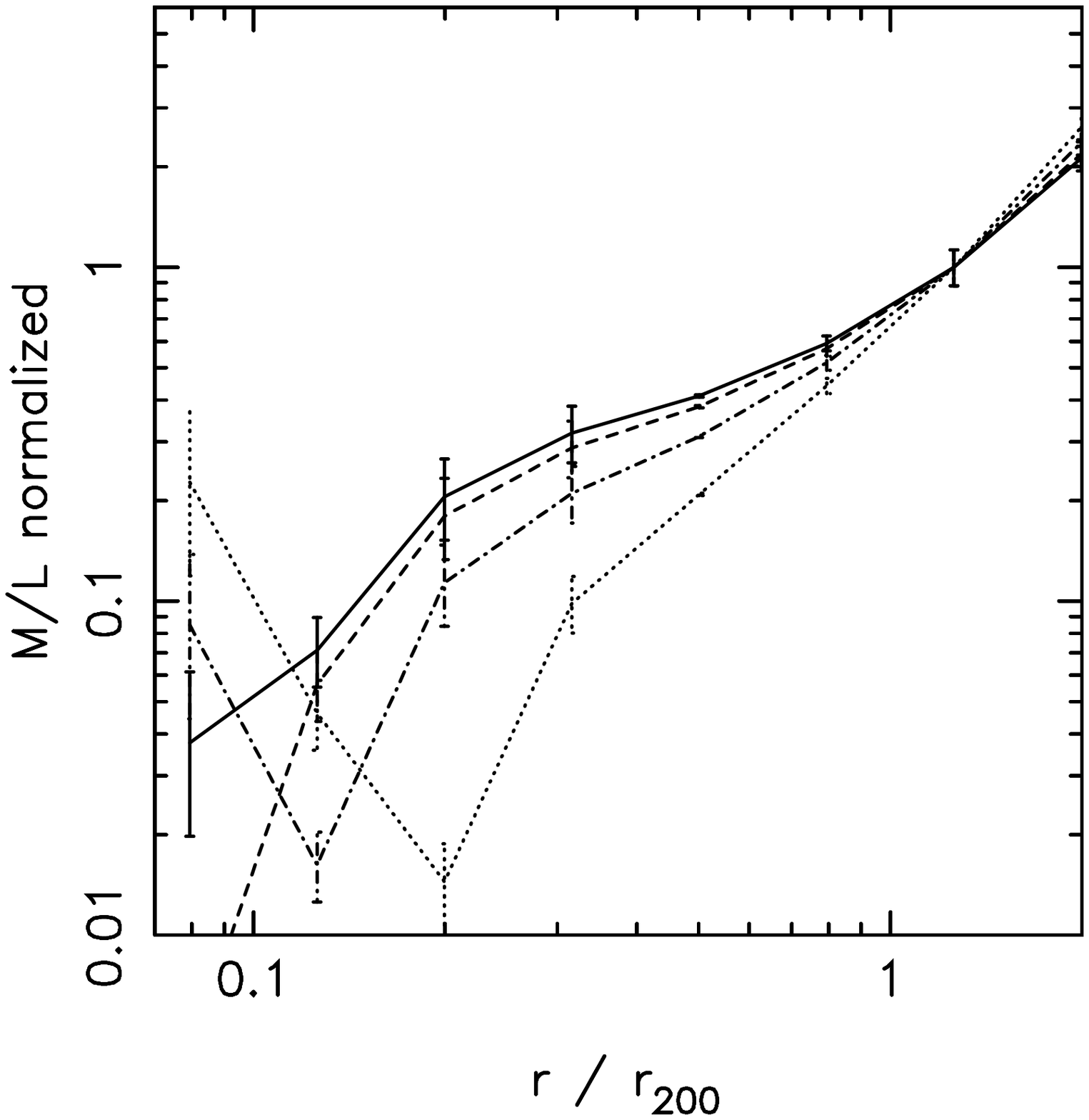}}
\end{picture}
\vspace{20.0cm}
\caption{}\end{figure}  

\begin{figure}\figurenum{\nfig}
\begin{picture}(0,0)(0,0)
        \put(  0,-550){\includegraphics[width=0.9\hsize]{fig15.ps} }
        \put( 80,-460){\includegraphics[width=0.3\hsize]{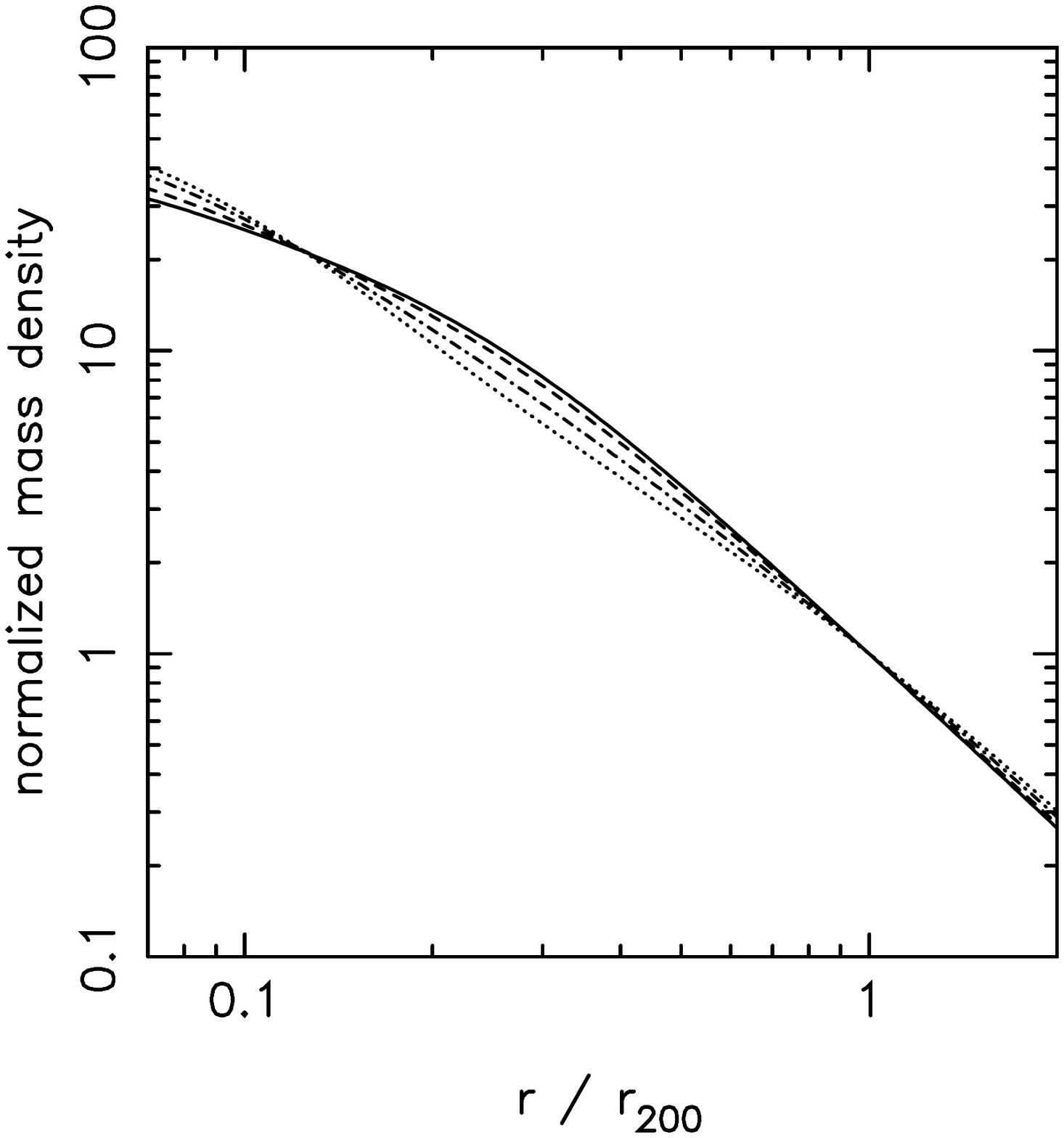}}
        \put(260,-290){\includegraphics[width=0.3\hsize]{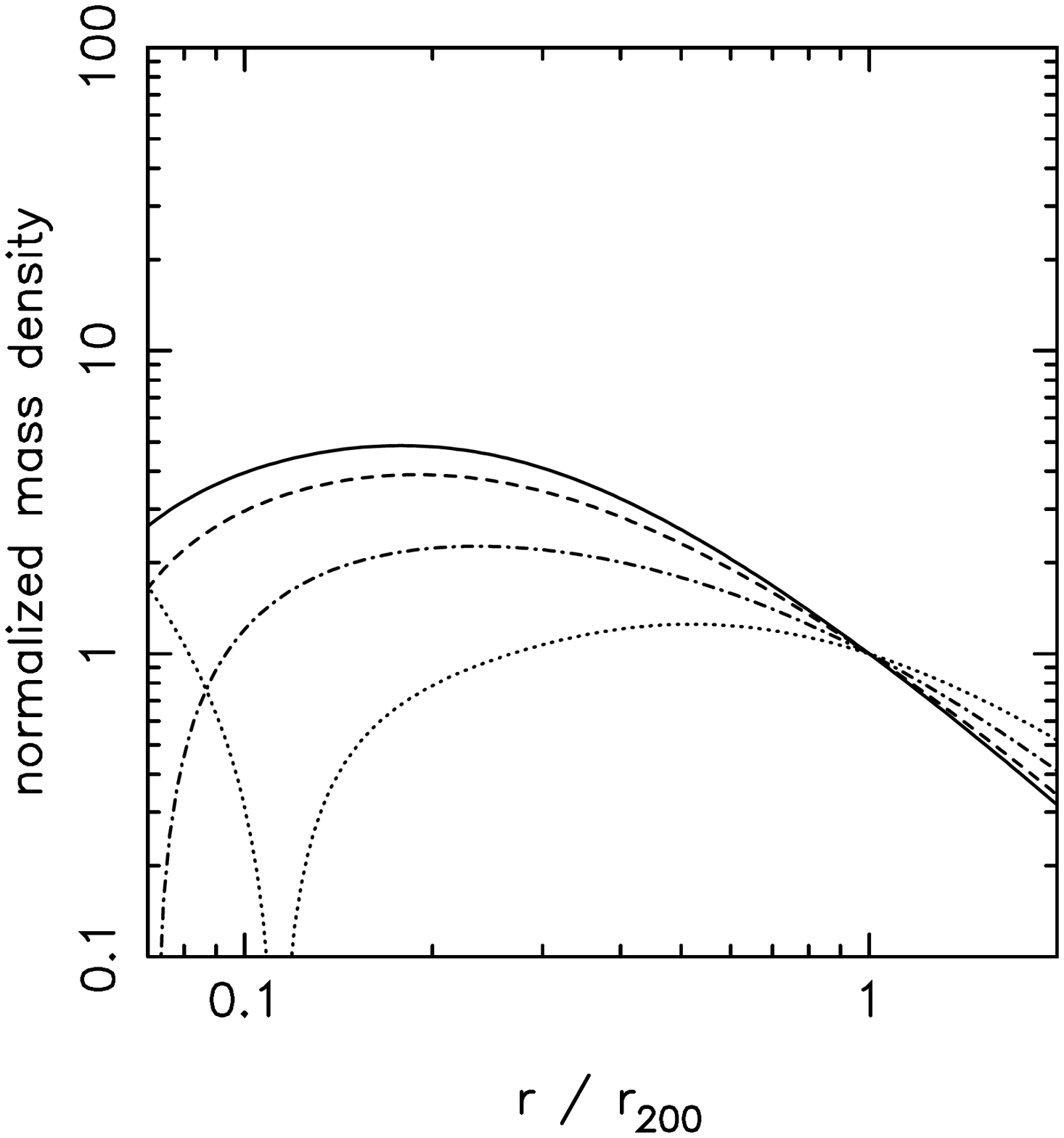}}
\end{picture}
\vspace{20.0cm}
\caption{}\end{figure}  

\begin{figure}\figurenum{\nfig}
	\includegraphics[width=0.9\hsize]{fig16.ps} 
\caption{}\end{figure}  
\begin{figure}\figurenum{\nfig}
	\includegraphics[width=0.9\hsize]{fig17.ps} 
\caption{}\end{figure}  

\begin{figure}\figurenum{\nfig}
	\includegraphics[width=0.9\hsize]{fig18.ps} 
\caption{}\end{figure}

\end{document}